\documentclass[11pt,a4paper]{article}
\pdfoutput=1
\usepackage{jheppub}

\def\beq{\begin{equation}}
\def\eq{\end{equation}}
\def\bea{\begin{eqnarray}}
\def\eea{\end{eqnarray}}

\def\nn{\nonumber}

\newcommand\fft[2]{\frac{#1}{#2}}
\newcommand\ft[2]{{\textstyle\frac{#1}{#2}}}

\newcommand{\be}{\begin{equation}} 
\newcommand{\ee}{\end{equation}} 
\newcommand{\barr}{\begin{array}}
\newcommand{\earr}{\end{array}}
\newcommand{\ea}{\end{eqnarray}}

\begin{document}

\preprint{LCTP-19-37}

\title{Higher-Derivative Corrections to Entropy and the\\ Weak Gravity Conjecture in Anti-de Sitter Space}

\author[a]{Sera Cremonini}
\emailAdd{cremonini@lehigh.edu}

\author[b]{Callum R. T. Jones}
\emailAdd{jonescal@umich.edu}

\author[b]{James T. Liu}
\emailAdd{jimliu@umich.edu}

\author[b]{Brian McPeak}
\emailAdd{bmcpeak@umich.edu}

\affiliation[a]{Department of Physics, Lehigh University, Bethlehem, PA, 18018}
\affiliation[b]{Leinweber Center for Theoretical Physics, Randall Laboratory of Physics\\ The University of Michigan, Ann Arbor, MI 48109-1040}

\abstract{We compute the four-derivative corrections to the geometry, extremality bound, and thermodynamic quantities of AdS-Reissner-Nordstr{\"o}m black holes for general dimensions and horizon geometries. We confirm the universal relationship between the extremality shift at fixed charge and the shift of the microcanonical entropy, and discuss the consequences of this relation for the Weak Gravity Conjecture in $\text{AdS}$. The thermodynamic corrections are calculated using two different methods: first by explicitly solving the higher-derivative equations of motion and second, by evaluating the higher-derivative Euclidean on-shell action on the leading-order solution. In both cases we find agreement, up to the addition of a Casimir energy in odd dimensions. We derive the bounds on the four-derivative Wilson coefficients implied by the conjectured positivity of the leading corrections to the microcanonical entropy of thermodynamically stable black holes. These include the requirement that the coefficient of Riemann-squared is positive, meaning that the positivity of the entropy shift is related to the condition that $c - a$ is positive in the dual CFT. We discuss implications for the deviation of $\eta/s$ from its universal value and a potential lower bound.}

\maketitle

\section{Introduction}

\noindent The aim of the swampland program \cite{Vafa:2005ui} is to understand what subset of the infinite space of possible effective field theories can arise at low energies from theories of quantum gravity. This requires finding simple criteria for classifying these theories; those that admit UV completions including quantum gravity are in the landscape, and those that do not are in the swampland. Such criteria are far more useful if they are detectable in the infrared without any knowledge of the particulars of the UV completion. In practice, one fruitful source of such IR intuition is the physics of black holes, which are believed to conform to general relativity and the laws of black hole thermodynamics in the semiclassical regime.

So far a number of criteria have been proposed (see e.g.~\cite{Palti:2019pca} for a review of the program). One that has attracted considerable interest is the Weak Gravity Conjecture (WGC), which roughly states that there are states whose mass is smaller than their charge \cite{ArkaniHamed:2006dz} --- hence, they are states for which ``gravity is the weakest force." Such states are labelled ``superextremal," and may be provided by fundamental particles or non-perturbative states such as black holes. The original motivation was to provide a mechanism for black holes to decay. These black holes are believed to obey an ``extremality bound" on its mass to charge ratio,
\begin{align}
    \frac{M}{Q} > 1 \, ,
\end{align}
which arises from requiring that the solution does not contain a naked singularity. It was believed that if a theory of gravity did not have a superextremal particle in its spectrum, then simple conservation of charge would prevent it from decaying to lighter components without violating cosmic censorship. This, in turn, would be problematic because it would lead to an infinite number of stable states or remnants. 

Another mechanism for the decay of nearly extremal black holes in flat space was pointed out in \cite{Kats:2007mq}. Generically, the low-energy limit of a theory of quantum gravity should include higher-derivative corrections that encode the UV physics in a highly suppressed way.
Such corrections can change the allowed charge to mass ratio of Reissner-Nordstr{\"o}m black holes so that they can decay to smaller black holes. In particular, the 
classical extremality bound would be corrected into a form that schematically may look like
\begin{align}
    \frac{M}{Q} > 1 - \frac{1}{Q^2} \left( 2 c_1 + 8 c_2 + ... \right) \, , 
\end{align}
shifted slightly by the Wilson coefficients $c_i$ of the higher-derivative operators. It then becomes possible for the nearly extremal black holes to decay as long as this combination of coefficients is positive. The statement that black holes can always decay through these higher-derivative corrections is sometimes called the ``Black Hole Weak Gravity Conjecture." This idea has inspired a large amount of work \cite{Cheung:2014vva, Cottrell:2016bty, Cheung:2018cwt, Andriolo:2018lvp, Hamada:2018dde, Charles:2019qqt, Jones:2019nev} on bounding the EFT coefficients, thereby proving the conjecture. While no existing proof is completely general, the work so far covers a large number of possibilities and assumptions. 

One intriguing proof of the WGC in flat space relates the extremality shift to the shift in the Wald entropy \cite{Cheung:2018cwt}. The authors first show that, near extremality, the shift to the extremality bound at fixed charge and temperature is proportional to the shift in entropy at fixed charge and mass. They then present an argument that the higher-derivative corrections should increase the entropy, thereby proving the Black Hole WGC. The argument for the entropy shift positivity is not expected to be fully general; it applies to higher-derivative corrections that arise from integrating out massive particles at tree-level. Nonetheless, it is curious that the entropy shift is proportional to the extremality shift. This fact was given a simple thermodynamic proof in \cite{Goon:2019faz}, where no assumptions were made about the particulars of the background.

So far, however, these ideas have not fully made their way to Anti-de Sitter space. From the WGC point of view, it is easy to see why: the relationship between mass and charge of an extremal black holes in AdS is \textit{already non-linear at the two-derivative level}%
\footnote{By ``extremal," we mean that the temperature is zero. This is not the same as the BPS limit in AdS.}.
Therefore it is not at all clear what is gained by studying the higher-derivative corrections to the extremal mass-to-charge ratio%
\footnote{Other aspects of the WGC have been discussed in AdS.  See e.g.~\cite{Nakayama:2015hga, Harlow:2015lma, Montero:2016tif, Montero:2018fns}.}.
Furthermore, massive particles emitted from a black hole cannot fly off to infinity in AdS as they can in flat space, so if the WGC allows for the instability of black holes in AdS, it must be through a completely different mechanism. (See the Discussion for more commentary on this possibility).

Regardless, the entropy-extremality relationship is expected to hold in AdS as it does in flat space (and indeed, an example in AdS${}_4$ was given in \cite{Goon:2019faz}). Therefore, this paper addresses two main issues in Anti-de Sitter space. First, we check the purported relationship between the entropy shift and the extremality shift, and indeed we find that
it holds for the AdS-Reissner-Nordstr{\"o}m backgrounds. Second, we examine the conjecture that the entropy shift is positive when the leading-order solution is a minimum of the action. By computing the entropy shift explicitly, we see that its positivity for stable black holes implies that the coefficient of Riemann-squared is \emph{universally positive}.
This has interesting consequences for the structure of $\eta/s$, as we comment on in section \ref{sec:discussion}.

This paper is organized as follows. In section \ref{sec:geo} we introduce the theory we will examine, and find the solutions to the equations of motion at first order in the EFT coefficients. In section \ref{sec:mce} we use the solution to compute the shift to extremality, considering the result both at fixed charge and fixed mass. We then compute the shift to the Wald entropy, and we find that the conjectured relationship of \cite{Cheung:2018cwt, Goon:2019faz} between the shift to mass and shift to entropy is valid for AdS-Reissner-Nordstr{\"o}m black holes. Furthermore, we notice that both the mass shift and entropy shift are also proportional to the charge shift, which allows us to extend the relationship to\footnote{This form of the entropy-extremality relation is valid away from extremality, but a slightly modified form is required for black holes which are very close to extremality. The subtleties involved in this limit are discussed in detail at appendix \ref{app:vnearext}.}  
\begin{align}
    \left(\Delta M \right)_{Q, T=0}= - T_0 \left(\Delta S \right)_{Q, M} = - \Phi_0 \left( \Delta Q \right)_{M, T = 0}\,.
\end{align}
We present a simple thermodynamic derivation of these relationships in appendix \ref{app:goonpenco}.

In section \ref{sec:euc}, we reproduce these results from a thermodynamic point of view. It has recently been shown \cite{Reall:2019sah} that the first-order corrections to the solutions are not needed to compute the first order corrections to thermodynamic quantities. In this section, we verify that this is the case for AdS-Reissner-Nordstr{\"o}m backgrounds by computing the four-derivative corrections to the renormalized on-shell action. From this we may compute the free energy and other thermodynamic quantities. We find that the results of this calculation match the results from section \ref{sec:mce} in even dimensions, while in odd dimensions the free energy and associated thermodynamic quantities are renormalization scheme dependent, and agree with the geometric calculation in a physically motivated \textit{zero Casimir} scheme.

In section \ref{sec:posent}, we first review the argument given in \cite{Cheung:2018cwt} for the positivity of the entropy shift, and comment on a potential issue with applying it to AdS. The positivity of the entropy shift requires that the black hole solutions are local minima of the path integral, so we compute the specific heat and electrical permittivity to determine the regions of parameter space where the black holes will be stable. Finally, we determine the constraints placed on the EFT coefficients by assuming that the entropy shift is positive for all stable black holes. We compare this to the results obtained by requiring that the entropy shift is positive for only extremal black holes, which is equivalent to the condition that the extremality shift at fixed charge and temperature is positive. The constraints include the requirement that the coefficient of Riemann-squared is positive. As this coefficient is proportional to the difference $c - a$ between the central charges of the dual CFT, we conclude that the positivity of the entropy shift will be violated in theories where $c - a < 0$.

We summarize our results in the section \ref{sec:discussion}, where we comment on the implications of our results for the behavior of $\eta/s$, as well as the nature of the WGC in AdS. We relegate to appendix \ref{app:entshift} the specific form of the entropy shifts and bounds on the EFT coefficients for AdS${}_5$ through AdS${}_7$. In appendix \ref{app:goonpenco}, we present a general proof of the entropy-extremality relation of \cite{Goon:2019faz}, and in appendix \ref{app:vnearext} we comment on some subtleties involving the extremal limit.

Note added: After completing this work, we noticed that the revised version of \cite{Goon:2019faz} obtains the same form of the entropy-extremality relation that we argue for in this paper.

\section{Corrections to the Geometry}
\label{sec:geo}

\noindent We consider Einstein-Maxwell theory in the presence of a negative cosmological constant in a $(d+1)$-dimensional
AdS spacetime of size $l$.
The first non-trivial terms in the derivative expansion of the effective action arise at the four-derivative level, and by appropriate field redefinitions we may choose a complete basis of dimension-independent operators:
\begin{align}
    \begin{split}
        I &= -\frac{1}{16 \pi } \int d^{d+1} x \sqrt{-g} \Bigg[  \frac{d (d-1)}{l^2} + R -\frac{1}{4} F^2 \\
        & \qquad \qquad \qquad \qquad +l^2  \epsilon \Big( c_1 R_{abcd} R^{abcd} +  c_2 R_{abcd} F^{ab} F^{cd} + c_3 (F^2)^2 + c_4 F^4 \Big) \Bigg].
    \end{split}
    \label{Action}
\end{align}
Note that additional CP-odd terms can arise in specific dimensions, but will not contribute to the static, stationary spherically symmetric black holes that we are considering here.  This basis parallels that of \cite{Myers:2009ij}, which used the same set of dimensionless Wilson coefficients, but focused on the $(4+1)$-dimensional case.  Depending on the origin of the AdS length scale $l$, one may expect these coefficients to be parametrically small, of the form $c_i\sim(\Lambda l)^{-2}$, where $\Lambda$ denotes the scale at which the EFT breaks down. In particular, this will be the case in order for the action (\ref{Action}) to be under perturbative control. We have also introduced the small bookkeeping parameter $\epsilon$, which will allow us to keep track of which terms are first order in the $c_i$ coefficients.

\subsection{The Zeroth Order Solution}

\noindent At the two-derivative level, this action admits a family of AdS-Reissner-Nordstr\"om black holes parametrized by uncorrected mass $m$ and charge $q$,
\begin{align}
    \begin{split}
        ds^2 =& -f(r) dt^2 + g(r)^{-1} dr^2 + r^2 d \Omega_{d - 1,k}^2 \, , \qquad f(r) = g(r) = k - \frac{m}{r^{d-2}} + \frac{q^2}{4 r^{2d-4}} + \frac{r^2}{l^2}, \\
        & \quad A = \left( - \frac{1}{c} \frac{q}{r^{d-2}} + \Phi \right) dt,  \qquad c = \sqrt{ \frac{ 2(d-2)}{(d-1)} }, \qquad \Phi = \frac{1}{c} \frac{q}{r_h^{d-2}}\, .
        \label{radial function}
    \end{split}
\end{align}
Here $r_h$ is the outer horizon radius, and the parameter $k=0,\pm1$ specifies the horizon geometry, with $k=1$ corresponding to the unit sphere.  The constant $\Phi$ is chosen so that the $A_t$ component of the gauge field vanishes on the horizon, and represents the potential difference between the asymptotic boundary and the horizon. 

Typically, we will consider lower case letters $(m, q, ...)$ to be parameters in the theory, while upper case letters $(M, Q, S, T, ...)$ will denote physical quantities that may or may not receive corrections. We will add a subscript zero (e.g. $M_0$) to denote the uncorrected contribution to quantities that do receive order $c_i$ corrections. The shifts, which are equal to the corrected quantities minus the uncorrected ones, will be denoted by the $\epsilon$ derivative. However, we will sometimes use $\Delta$ when it is convenient, with subscripts indicating quantities held fixed, for example, we have 
\begin{align}
    (\Delta M)_T \equiv \lim_{\epsilon \rightarrow 0}\left( M(T,\epsilon) - M_0(T)\right)  \equiv \lim_{\epsilon \rightarrow 0}\left( \frac{ \partial M}{\partial \epsilon} \right)_T. 
\end{align}
Finally, in sections \ref{sec:euc} and \ref{sec:posent} we will use dimensionless quantities $(\nu, \xi)$ for convenience. These are defined by $\nu = (r_h)_0 / l$ and $Q = (1 - \xi) Q_{\text{ext}} $.

\subsection{The First Order Solution}

\noindent We now turn to the first order solution in terms of the Wilson coefficients $c_i$.  We follow the procedure outlined in Ref.~\cite{Kats:2006xp}, but work in an AdS$_{d+1}$ background.  While general $(d+1)$-dimensional results may be worked out analytically, we took a shortcut of working with explicit dimensions four through eight and then fitting the coefficients to extract results for arbitrary dimension.  Since the four-derivative terms are built from tensors with eight indices and hence four metric contractions, the resulting expressions will scale at most as $d^4$.  The coefficients are hence fully determined by working in five different dimensions.

Following \cite{Kats:2006xp}, we start with the \textit{effective} stress tensor, where corrections come from two sources. The first is from substituting in the corrected Maxwell field to the zeroth order electromagnetic stress tensor, and the second is from the explicit four-derivative corrections to the stress tensor evaluated on the zeroth order solution.  The result of computing both of these contributions to the time-time component of the stress tensor is
\begin{align}
    \begin{split}
        T_t{}^t&= -\frac{(d-1) (d-2) \, q^2}{4 \, r^{2d - 2}} 
        + \frac{d (d-1)}{ l^2} \\
        & + c_1 \Bigg( \frac{ (d - 2) (8 d^3 - 24 d^2 + 15 d + 3) \, q^4 l^2}{ 8 r^{4d - 4}}
        - \frac{ (d - 1) (d - 2) (4 d^2 - 9 d + 3 ) \, m q^2 l^2 }{r^{3d - 2}} 
        \\
        & \qquad \qquad \qquad + k \frac{4  d (d-1) (d-2)^2 \, l^2 q^2}{r^{2d}}
        - \frac{ d (d-1) (d-2) (d-3) \, l^2 m^2}{r^{2d}} \\
        & \qquad \qquad \qquad + \frac{(d- 2) (2d - 3)(2 d^2 - 5 d + 1) \, q^2 }{r^{2 d-2}} + \frac{2 d (d - 3) }{l^2} \Bigg)
        \\
        & + c_2 \Bigg( \frac{ (d-1)^3 (d-2) \, q^4 l^2}{r^{4d-4}}
        - \frac{(d-1)^2 (3d^2 - 8 d + 4) \, q^2 m l^2}{r^{3d-2}} 
        + k \frac{2 d (d-1)^2 (d-2) \, q^2 l^2}{r^{2d}} \\
        & \qquad \qquad \qquad + \frac{2 (d-1)^3 (d-2) \, q^2}{r^{2d-2}} \Bigg)
         + \left( 2 c_3 + c_4 \right) \Bigg( \frac{(d-1)^2 (d-2)^2 q^4 l^2}{2 \, r^{4 d-4}} \Bigg) \, .
         \label{StressTensor}
    \end{split}
\end{align}
The shift to the geometry may be obtained from the corrections to the stress tensor \cite{Kats:2006xp}, 
\begin{align}
    \Delta g = \frac{1}{(d - 1) r^{d - 2}} \int \, dr \,  r^{d-1}\Delta T_t{}^t \, ,
\end{align}
and after integrating the $\mathcal{O}(c_i)$ terms in (\ref{StressTensor}), we find 
\begin{align}
    \begin{split}
        &\Delta g (r) = \\ 
        &c_1 \Bigg( -\frac{ (d - 2) (8 d^3 - 24 d^2 + 15 d + 3) \, q^4 l^2}{ 8 (d - 1) (3 d - 4) r^{4d - 6}}
        + \frac{(d - 2) (4 d^2 - 9 d + 3 ) \, m q^2 l^2 }{2 (d - 1) r^{3d - 4}} 
        \\
        & \qquad  - k \frac{4  (d-2)^2 \, l^2 q^2}{r^{2d-2}}
        + \frac{  (d-2) (d-3) \, l^2 m^2}{r^{2d-2}}
        - \frac{ (2d - 3)(2 d^2 - 5 d + 1) \, q^2 }{(d - 1)r^{2 d-4}}
        + \frac{2 (d - 3) r^2}{(d - 1)l^2} \Bigg)
        \\
        & + c_2 \Bigg( - \frac{ (d-1)^2 (d-2) \, q^4 l^2}{(3d - 4)r^{4d - 6}}
        + \frac{ (3d^2 - 8 d + 4) \, q^2 m l^2}{2 r^{3d-4}}
        - k \frac{2 (d-1) (d-2) \, q^2 l^2}{r^{2d-2}} \\
        &\hspace{12mm}- \frac{2 (d-1)^2 \, q^2}{r^{2d-4}} \Bigg) \\
        & + \left( 2 c_3 + c_4 \right) \Bigg(- \frac{(d - 1) (d - 2)^2 q^4 l^2 }{ (6d - 8) r^{4d - 6}} \Bigg) \, .
    \end{split}
    \label{eq:Deltag}
\end{align}
The time component of the metric can then be obtained using the relation \cite{Kats:2006xp} 
\begin{align}
    f(r) = (1 + \gamma(r)) g(r),
\end{align}
where $\gamma(r)$ is defined by\footnote{
We note that the definition of $\gamma$ implies that it is positive provided that the null energy condition holds. 
}
\begin{align}
    \gamma(r) =    -\frac{1}{(d - 2)} \int dr r \left( T_t{}^t - T_r{}^r \right) .
    \label{eq:gamma}
\end{align}
For our particular case we find:
\begin{align}
    \gamma(r) =\left( c_1 \frac{(d - 2)(2 d^2 - 5 d + 1) }{(d - 1) } + c_2 d (d - 2)\right)\fft{q^2l^2}{r^{2d-2}}.
    \label{eq: gammaVal}
\end{align}
Finally, we have
\begin{align}
    \begin{split}
        F_{tr} &= \sqrt{\frac{(d - 2)(d - 1) }{2}} \Bigg[(1-8c_2) \frac{q}{r^{d - 1}}+4c_2(d-1)(d-2)\fft{qml^2}{r^{2d-1}}\\
        &\kern4em+\left( \fft{c_1}2\frac{(2 d^2 - 5 d + 1)}{(d - 1)} -\fft{ c_2}2 (7d - 12) - 4\left( 2 c_3 + c_4 \right)(d - 1) \right) (d-2)\fft{q^3l^2}{r^{3d-3}}
        \Bigg] \, ,
    \end{split}
    \label{eq:Ftr}
\end{align}
which we note is independent of the geometry parameter $k$, as was the case in \cite{Cremonini:2009ih}.

\subsection{Asymptotic Conditions and Conserved Quantities}

\noindent The first order solution can be summarized as
\begin{align}
    &ds^2=-\left(1+\gamma(r)\right)g(r)dt^2+g(r)^{-1}dr^2+r^2d\Omega_{d-1,k}^2 \, ,
\end{align}
where
\begin{equation}
    g(r)=k-\fft{m}{r^{d-2}}+\fft{q^2}{4r^{2d-4}}+\fft{r^2}{l^2}+\Delta g.
\end{equation}
The corrected metric functions, $\Delta g$ and $\gamma(r)$, are given in (\ref{eq:Deltag}) and (\ref{eq:gamma}), respectively.  In addition, the full electric field is given in (\ref{eq:Ftr}).  For a given zeroth order AdS radius $l$, this solution is specified by two parameters, $m$ and $q$, which correspond to the mass and charge of the uncorrected black hole.
At the same time, the corrected solution includes a number of integration constants, two of which we have implicitly set to zero in the integral expressions for $\Delta g$ and $\gamma(r)$.  The constant related to $\Delta g$ can be absorbed by a shift in $m$, and a third constant from the corrected Maxwell equation can be absorbed by a shift in $q$.  The constant related to $\gamma(r)$ can be absorbed at the linearized level by a rescaling of the time coordinate, and hence can be thought of as a redshift factor.

In order to make the correspondence between the parameters of the solution, $m$ and $q$, and the physical mass and charge of the black hole more precise, consider the part of $\Delta g$ that is leading in $r$. We can see that there is a term that goes like $c_1 \frac{r^2}{l^2}$ that dominates over all other terms in the correction. Therefore, for large values of $r$, the solution takes the form
\begin{align}
    f(r)\approx g(r)&=k-\fft{m}{r^{d-2}}+\left(1+c_1\fft{2(d-3)}{d-1}\right)\fft{r^2}{l^2}+\cdots,\nn\\
    F_{tr}&=\sqrt{\fft{(d-2)(d-1)}2}(1-8c_2)\fft{q}{r^{d-1}}+\cdots.
\end{align}
Our first observation is that the AdS radius gets modified because the Riemann-squared term is non-vanishing on the original uncorrected background.  This suggests that we define an effective AdS radius
\begin{align}
    l^2 = \lambda^2 l^2_{\text{eff}} , \qquad \qquad \lambda^2 = \left( 1 + c_1 \frac{2 (d - 3)}{(d - 1)} \right).
    \label{eq:lamdef}
\end{align}
This shift by $\lambda$ is unavoidable when turning on the $c_1$ Wilson coefficient.  However, in principle we still have a choice of whether we hold $l$ or $l_{\text{eff}}$ fixed when turning on the four-derivative corrections.

In what follows, we always choose to keep $l$ fixed.  Then, since the effective AdS radius is shifted, the asymptotic form of the metric is necessarily modified as well.  From a holographic point of view, this leads to a modification of the boundary metric
\begin{equation}
    ds^2\sim r^2\left(\fft{dt^2}{l^2}+d\Omega_{d-1,k}^2\right)\quad\longrightarrow\quad ds^2\sim r^2\left(\fft{dt^2}{l_{\text{eff}}^2}+d\Omega_{d-1,k}^2\right).
\end{equation}
This is generally undesirable, as we would like to compare thermodynamic quantities in a framework where we hold the boundary metric fixed while turning on the Wilson coefficients.  One way to avoid this shift in the boundary metric is to introduce a `redshift' factor
\begin{equation}
    t=\bar t/\lambda,
\end{equation}
to compensate for the shift in $l_{\text{eff}}$.  In terms of the time $\bar t$, the solution now takes the form
\begin{align}
    ds^2&=-\bar f(r)\, d\bar t^2+g(r)^{-1}dr^2+r^2d\Omega_{d-1,k}^2,\nn\\
    F_{\bar tr}&=\lambda^{-1}F_{tr}=\sqrt{\fft{(d-2)(d-1)}2}(1-8c_2)\fft{q/\lambda}{r^{d-1}}+\cdots,
\end{align}
where
\begin{align}
    \bar f(r)&=\lambda^{-2}(1+\gamma(r))g(r)=k/\lambda^2-\fft{m/\lambda^2}{r^{d-2}}+\fft{r^2}{l^2}+\cdots,\nn\\ g(r)&=k-\fft{m}{r^{d-2}}+\fft{r^2}{l_{\text{eff}}^2}+\cdots.
\end{align}

We now turn to the charge and mass of the solution measured with respect to the redshifted $\bar t$ time.  For the charge $Q$, we take the conserved Noether charge
\begin{equation}
    Q=\fft1{16\pi}\int_{\Sigma_{d-1}}*\mathcal F,
\end{equation}
where $\mathcal F$ is the effective electric field
\begin{equation}
    \mathcal F_{\mu\nu}=F_{\mu\nu}+l^2\left(-4c_2R_{\mu\nu\rho\sigma}F^{\rho\sigma}-8c_3F_{\mu\nu}(F^2)-8c_4F_{\nu\rho}F^{\rho\sigma}F_{\sigma\mu}\right).
\end{equation}
The result is
\begin{equation}
    Q=\left.\fft{1+8c_2}{16\pi}\omega_{d-1}\lambda r^{d-1}F_{\bar t r}\right|_{r\to\infty}=\sqrt{\fft{(d-2)(d-1)}2}\fft{\omega_{d-1}}{16\pi}q,
    \label{eq:Qdef}
\end{equation}
where $\omega_{d-1}$ is the volume of the unit $S^{d-1}$.  The $1/16\pi$ factor arises from the prefactor in the action (\ref{Action}) where we have set Newton's constant $G=1$.

Unlike in the asymptotically Minkowski case, some care needs to be taken in obtaining the mass of the black hole.  With an eye towards holography, we choose to define the mass from the boundary stress tensor \cite{Balasubramanian:1999re}.  The standard approach to holographic renormalization involves the addition of appropriate local boundary counterterms so as to render the action finite.  This was performed in \cite{Cremonini:2009ih} for $R^2$-corrected bulk actions, and since only the $c_1R_{abcd}R^{abcd}$ term in (\ref{Action}) leads to an additional divergence, we can directly use the result of \cite{Cremonini:2009ih}.  The result is
\begin{equation}
    M=\fft{\omega_{d-1}}{16\pi}(1+4c_1(d-3))\fft{(d-1)m}\lambda,
\end{equation}
where we have taken into account the scaling of the mass by the redshift factor $\lambda$.  Substituting in $\lambda$ from (\ref{eq:lamdef}) then gives
\begin{equation}
    M=\fft{\omega_{d-1}}{16\pi}(d-1)(1+\rho)m,
    \label{eq:massdef}
\end{equation}
where
\begin{align}
    \rho = c_1 \frac{(d - 3) (4 d - 5)}{d - 1}.
    \label{eq:rhodef}
\end{align}
Note that we are taking the mass here to \textit{exclude} the Casimir energy that is normally part of the boundary stress tensor.  This will be important when comparing with the thermodynamic quantities extracted from the regulated on-shell action in section \ref{sec:euc}.  Working in the setup of holographic renormalization ensures that the mass $M$ and charge $Q$ defined in (\ref{eq:massdef}) and (\ref{eq:Qdef}), respectively, yield a consistent framework for black hole thermodynamics.  

\section{Mass, Charge, and Entropy from the Geometry Shift}
\label{sec:mce}

\noindent Given the first-order solution, we now consider shifts to the mass, $\Delta M$, and entropy, $\Delta S$, of the black hole induced by the four-derivative corrections. In these computations it is important to keep in mind what is being held fixed as we turn on the Wilson coefficients $c_i$.  The main parameters we consider here are the mass $M$ and charge $Q$, which are related to the two parameters, $m$ and $q$, of the solution by (\ref{eq:massdef}) and (\ref{eq:Qdef}), respectively. In addition we consider the thermodynamic quantities $T$ (temperature) and $S$ (entropy), although they are not all independent.  Note that we always consider the AdS radius $l$ to be fixed, although interesting results have been obtained by mapping it to thermodynamic pressure.

Singly-charged, non-rotating black holes may be described by any two of mass $M$, charge $Q$ and the horizon radius $r_h$.  Of course, any number of other parameters may be used as well, such as the temperature $T$ or an extremality parameter, such as was used in \cite{Cheung:2018cwt}. If we further impose the extremality condition $T=0$ on the solution, then only a single parameter is needed. Clearly this is only true for non-rotating black holes with a single gauge field, as more general solutions may have additional charges or angular momenta.

Here we mainly focus on the effect of higher-derivative corrections on extremal or near extremal black holes.  In particular, we consider the extremality shift $\Delta(M/Q)$ and the entropy shift $\Delta S$.  However, it is important to keep in mind what is being held fixed when we turn on the higher-derivative corrections, as the results will depend on this choice.  For example, we will see below that the shift to $M/Q$ depends on whether the mass, charge or horizon radius is held fixed when comparing the corrected with uncorrected quantities. 

\subsection{Mass, Charge, and Extremality}

\noindent Recall that, in our first-order solution, the geometry is essentially given by the radial function
\begin{align}
\label{gcorr}
    g^{rr}=g(r) = k - \frac{m}{r^{d-2}} + \frac{q^2}{4 r^{2d-4}} + \frac{r^2}{l^2} + \, \Delta g \, ,
\end{align}
where $\Delta g$ denotes the contributions of the higher-derivative corrections to the geometry, and $\epsilon$ is a small parameter we use to keep track of where $\mathcal{O}(c_i)$ corrections come in. Using the fact that both $g(r_h)$ and $g'(r_h)$ vanish at extremality, 
we may express the extremal mass and charge as a function of the horizon radius,

\begin{align}
    \begin{split}
        M_{\text{ext}} &= 2 V (d - 1) r_h^{d-2} \left( \left(k +  \frac{d - 1}{d - 2 } \frac{r_h^2}{l^2} \right) \left(1 + \epsilon \rho \right) + \epsilon \, \Delta g + \frac{r_h}{2(d - 2)} \epsilon \, \Delta g' \right) \, , \\
        Q_{\text{ext}}^2 & = 2 V^2 (d - 1)(d - 2) r_h^{2(d-2)} \left( k + \frac{d }{d - 2 } \frac{r_h^2}{l^2}  + \epsilon \, \Delta g + \frac{r_h}{d - 2} \epsilon \, \Delta g' \right) \, ,
        \label{extremal_mq}
    \end{split}
\end{align}
where $M$ and $Q$ are the asymptotic quantities defined in (\ref{eq:massdef}) and (\ref{eq:Qdef}), and we have defined $V=\omega_{d-1}/16\pi$. Though we have expressed $M$ and $Q$ as functions of $r_h$, these expressions are valid regardless of which of the three quantities is being held fixed. For example, if we work at fixed charge, then $Q$ gets no $\mathcal{O}(\epsilon)$ corrections, in which case $M$ and $r_h$ will both receive corrections.  

\subsubsection{Extremality at Leading Order}

\noindent Before discussing the extremality and entropy shifts, we consider the leading order relations between $M_0$, $Q_0$ and $(r_h)_0$ for extremal black holes. We will suppress the $0$ subscripts in this subsection, but we mean the uncorrected quantities. Setting $\epsilon=0$ in (\ref{extremal_mq}) immediately gives the relations
\begin{align}
    \begin{split}
        M_{\text{ext}} &= 2 V (d - 1) r_h^{d-2} \left(k +  \frac{d - 1}{d - 2 } \frac{r_h^2}{l^2} \right)\, ,\\
        Q_{\text{ext}}^2 &= 2 V^2 (d - 1)(d - 2) r_h^{2(d-2)} \left( k + \frac{d }{d - 2 } \frac{r_h^2}{l^2} \right) \, .
        \label{leading_extremal_mq}
    \end{split}
\end{align}
In principle, we can eliminate $r_h$ from these equations to obtain the relation between mass and charge for extremal AdS black holes.  However, for general dimension $d$, there is no simple expression that directly encodes this relation.  Nevertheless, we can consider the limit of small and large black holes.

For small black holes ($r_h \ll l$), we take $k=1$ (i.e. a spherical horizon) and find
\begin{align}
    M_{\text{ext}} \sim Q_{\text{ext}} \sim r_h^{d - 2} \, ,
\end{align}
so one recovers the simple $M \sim Q$ scaling that appears in flat space.  (Note that asymptotically Minkowski black holes necessarily have spherical horizons.)  For large black holes ($r_h \gg l$), on the other hand, the scaling is very different from that of flat space,
\begin{align}
    M_{\text{ext}} \sim r_h^{d} \, , \qquad Q_{\text{ext}} \sim r_h^{d - 1}\qquad\Rightarrow\qquad M_{\text{ext}} \sim \left( Q_{\text{ext}} \right)^{\frac{d}{d - 1}} \, .
\end{align}
In fact, this is precisely the scaling behavior expected based on the relationship between minimal scaling dimension and charge for boundary operators with large global charges \cite{Loukas:2018zjh}.

\subsubsection{Mass Shift at Fixed Charge}

\noindent Now we consider the effect of four-derivative corrections. If we hold the charge fixed, then the shift to extremality is entirely due to the change in the mass.  This may computed from the expression (\ref{extremal_mq}) for the mass by taking a derivative with respect to $\epsilon$, which parametrizes the higher-derivative corrections, leading to
\begin{align}
    \begin{split}
     & \left( \frac{ \partial M }{\partial \epsilon} \right)_{Q, T = 0} =  V (d - 1) r_h^{d - 2} \Bigg( 2 \Delta g + \frac{1}{d - 2} r_h \Delta g' \\
     & \quad + 2 \rho \left( k + \frac{d - 1}{d - 2} \frac{r_h^2}{l^2} \right) + \frac{2}{(d - 2) r_h} \left( (d - 2)^2 k + d (d - 1) \frac{r_h^2}{l^2}  \right) \left( \frac{ \partial r_h } {\partial \epsilon} \right)  \Bigg) \, ,
    \end{split}
\end{align}
where we have taken into account the fact that when the charge is fixed, we must allow the horizon radius $r_h$ to vary with $\epsilon$. To compute the shift $\partial r_h/\partial\epsilon$, we use the fact that we are holding $Q$ fixed. Then we use the expression for $Q_{\text{ext}}$ in (\ref{extremal_mq}) and demand that $\left({\partial Q}/{\partial \epsilon} \right)_{T = 0}= 0$ to obtain an equation for $\partial r_h/\partial\epsilon$. This procedure leads to the rather simple result
\begin{align}
    \begin{split}
     & \left( \frac{ \partial M } {\partial \epsilon} \right)_{Q, T = 0} = V (d - 1)  r_h^{d - 2} \left( \Delta g + 2 \rho \left( k + \frac{d - 1}{d - 2} \frac{r_h^2}{l^2} \right) \right) \, .
     \label{mass_shift}
    \end{split}
\end{align}
Note that the dependence on $\Delta g'$ has vanished. From the geometric point of view, this non-trivial cancellation is crucial for the extremality-entropy relation to hold. 

\subsubsection{Charge Shift at Fixed Mass}

\noindent If we instead hold the mass fixed, the entire shift in the extremality is due to the shift in charge. Following the same procedure as in the fixed charge case, but this time demanding $\partial M_{\text{ext}}/\partial\epsilon=0$, we find the relation:
\begin{align}
    \left( \frac{ \partial Q^2 } {\partial \epsilon} \right)_{M, T = 0} = - 2 V^2 (d - 1) (d - 2) r_h^{2d - 4} \left( \Delta g + 2 \rho  \left( k + \frac{d - 1}{d - 2} \frac{r_h^2}{l^2}\right) \right) \, .
\end{align}
Here we also find a cancellation of all $\Delta g'$ terms. Moreover, this shift is proportional to the mass shift at fixed charge
\begin{equation}
     \left( \frac{ \partial Q^2 } {\partial \epsilon} \right)_{M, T = 0}=-2V(d-2)r_h^{d-2}\left( \frac{ d M } {d \epsilon} \right)_{Q, T = 0}\,.
\end{equation}
This relationship more clear when we write this as the shift of $Q$ rather than $Q^2$. Using $\Delta Q^2 = 2 Q \Delta Q$, we find
\begin{align}
        Q \left( \frac{ \partial Q } {\partial \epsilon} \right)_{M, T = 0}=- V(d-2)r_h^{d-2}\left( \frac{ \partial M } {\partial \epsilon} \right)_{Q, T = 0}\,.
\end{align}
Finally, we use $\Phi = Q / (d - 2)  V  r^{d - 2}$ to write:
\begin{align}
        \left( \frac{ \partial M } {\partial \epsilon} \right)_{Q, T = 0} = 
        - \Phi \left( \frac{ \partial Q } {\partial \epsilon} \right)_{M, T = 0} \,.
\end{align}
So we see that the mass shift is related to the charge shift times the potential. In appendix \ref{app:goonpenco}, we derive this statement for a general thermodynamic system and show that it holds for any extensive charge and its conjugate. 

One physical consequence of this fact is that the entropy-extremality relationship (with a different proportionality factor) will hold regardless of whether the mass or charge is held fixed. As far as we know, this has not been noticed before in the literature.

\subsubsection{Summary of Extremality Shifts}

\noindent The shifts to extremality may be obtained from these mass and charge shifts. For completeness, we also present calculation at fixed horizon radius, as this extremality shift has previously been considered in the literature as well \cite{Myers:2009ij, Cremonini:2009ih},
\begin{align}
    \begin{split}
        \left( \frac{M}{Q} \right)_{Q, T = 0} \quad &= \quad  \left( \frac{M}{Q}  \right)_0  \left( 1 + \rho +  \Delta g \ \frac{1}{2\left( k + \frac{d - 1}{d - 2} \frac{r_h^2}{l^2} \right)} \right), \\
        \left( \frac{M}{Q} \right)_{M, T = 0} \quad &= \quad  \left( \frac{M}{Q}  \right)_0  \left( 1 + \rho \ \frac{k + \frac{d - 1}{d - 2} \frac{r_h^2}{l^2}}{k + \frac{d}{d - 2} \frac{r_h^2}{l^2}} +  \Delta g \  \frac{1}{2\left( k + \frac{d}{d - 2} \frac{r_h^2}{l^2} \right)}   \right), \\ 
        \left( \frac{M}{Q} \right)_{r_h, T = 0} \quad &= \quad \left( \frac{M}{Q}  \right)_0 \left(  1 + \rho + \frac{ \Delta g \left( k + \frac{d + 1}{d - 2} \frac{r_h^2}{l^2}  \right) + r_h \Delta g' \frac{1}{(d - 2)^2} \frac{r_h^2}{l^2}} {2 \left( k + \frac{d - 1}{d - 2} \frac{r_h^2}{l^2}  \right) \left( k + \frac{d}{d - 2} \frac{r_h^2}{l^2}  \right) } \right),
    \end{split}
    \label{eq:MQ3}
\end{align}
where the corrections are encoded in $\rho$ and $\Delta g$ given in (\ref{eq:rhodef}) and (\ref{eq:Deltag}), respectively (and $\Delta g'$ as well for the fixed $r_h$ case).  For these final results, we have set $\epsilon=1$.  However, the expressions are only valid to first order in the Wilson coefficients $c_i$.  Here the uncorrected charge to mass ratio may be obtained from (\ref{leading_extremal_mq}), and takes the form
\begin{align}
    \left( \frac{M}{Q}  \right)_0 \quad = \quad   \sqrt{\frac{2 (d - 1)}{d - 2}} \frac{ k +  \frac{d - 1}{d - 2 } \frac{r_h^2}{l^2} }{  \sqrt{k + \frac{d }{d - 2 } \frac{r_h^2}{l^2}} } \, .
    \label{eq:MQ0}
\end{align}
Note that, in (\ref{eq:MQ3}), the horizon radius $r_h$ may be taken to be the uncorrected radius, and can be obtained from either $M$ or $Q$ using the leading order expressions (\ref{leading_extremal_mq}).  In (\ref{eq:MQ0}), the leading order expression for $r_h$ should be used.  Finally, note that $\Delta g$ depends on the parameters $m$ and $q$ as well as the radius $r$.  The $m$ and $q$ parameters are directly obtained from $M$ and $Q$ using (\ref{eq:massdef}) and (\ref{eq:Qdef}), and again the leading order horizon radius can be used in $\Delta g$.

\subsection{Wald Entropy}
\label{sec:wald}

\noindent We now compare the shift in mass at fixed charge and temperature to the shift in entropy at fixed mass and charge. The entropy for black holes in higher-derivative theories is given by the Wald entropy \cite{Wald:1993nt}:
\begin{align}
    S = - 2 \pi \int_{\Sigma} \frac{\delta \mathcal{L}}{\delta R_{\mu \nu \rho \sigma}} \epsilon_{\mu \nu} \epsilon_{\rho \sigma} \, .
    \label{Wald_Entropy}
\end{align}
For spherically symmetric backgrounds, the integral over the horizon $\Sigma$ gives a factor of the area $A$. The two-derivative contribution to the entropy is simply $S^{(2)}=A/4$, while the four-derivative terms yield
\begin{align}
    S^{(4)} =\left. -2\pi A \frac{\delta \Delta \mathcal{L}}{\delta R_{\mu\nu\rho\sigma}} \epsilon_{\mu\nu}\epsilon_{\mu\nu}\right|_{\partial^4} = - \frac{A}{4} l^2 (4 c_1 R_{trtr} + 2 c_2 F_{tr}F_{tr}) \, .
\end{align}
The total entropy is the sum of these terms,
\begin{align}
    \begin{split}
        S =\left. \frac{A}{4} \left(1 - \epsilon  \left( 4 c_1 l^2 R_{trtr} + 2 c_2 l^2 F_{tr} F_{tr} \right) \right)\right|_{r_h} \,,
    \end{split}
    \label{eq:Wald}
\end{align}
where we once again introduced $\epsilon$ to parametrize the expansion.  Here the horizon area is given by $A=\omega_{d-1}r_h^{d-1}$, where $r_h$ is the corrected horizon radius.  On the other hand, the $R_{trtr}$ and $F_{tr}F_{tr}$ terms need only be computed on the zeroth-order background,
\begin{align}
    \begin{split}
        R_{trtr} &= \frac{1}{l^2} + \frac{(2 d - 3)(Q/V)^2}{2 (d - 1) r^{2 d - 2}} - \frac{(d - 2) M/V}{2 r^{d}} \, , \\
        F_{tr}F_{tr} &= \frac{(Q/V)^2}{r^{2 d - 2}} \, .    
    \end{split}
\end{align}
It does not matter whether we use the corrected or uncorrected quantities here because they already show up in a term that is order $\epsilon$. Note also that, while the expression for the Wald entropy (\ref{eq:Wald}) is given in terms of $M$, $Q$ and $r_h$ of the fully corrected solution, only two of these quantities are independent.

We now examine the entropy shift for a given solution at fixed mass $M$ and charge $Q$.  For the moment, we work at arbitrary $M$ and $Q$, and not necessarily at extremality.  The general expression for the entropy shift is then
\begin{equation}
    \left(  \frac{\partial S}{\partial \epsilon} \right)_{Q, M} = \frac{A}{4} \left( (d - 1)\left( \frac{1}{r_h} \frac{ \partial r_h }{\partial \epsilon}\right)_{Q,M}  -  \left( 4 c_1 l^2 R_{trtr} + 2 c_2 l^2 F_{tr} F_{tr} \right) \right) \, ,
    \label{eq:Sshift}
\end{equation}
where the first term was obtained by
\begin{align}
    \frac{1}{A} \frac{ \partial A }{\partial \epsilon} = (d - 1) \frac{1}{r_h} \frac{ \partial r_h }{\partial \epsilon} \, .
\end{align}
Here, it is important to note that the horizon radius $r_h$ receives a correction when working at fixed $M$ and $Q$.  If, on the other hand, we were to keep the horizon radius fixed (as is done in \cite{Myers:2009ij}), we would find only the second (interaction) term in (\ref{eq:Sshift}), and the entropy shift would be independent of $c_3$ and $c_4$.

To compute ${\partial r_h}/{\partial \epsilon}$, we start with the horizon condition $g(r_h)=0$ where $g(r)$ is given by (\ref{gcorr}) with $m$ and $q$ rewritten in terms of $M$ and $Q$.  Taking a derivative and solving for $\partial r_h/\partial\epsilon$ then gives
\begin{align}
    \fft1{r_h}\frac{ \partial r_h }{\partial \epsilon} =  - \frac{\rho M+V(d-1)r_h^{d-2}\Delta g} {(d-2)(M-(M_{\text{ext}})_0)} \, .
    \label{eq:drhde}
\end{align}
where $(M_{\text{ext}})_0$ is the leading order extremal mass given in (\ref{leading_extremal_mq}).  As we can see, this expression diverges if the leading order solution is extremal.  This is in fact not a surprise, as leading order extremality implies a double root at the horizon.  The higher order corrections will lift this double root and hence cannot be parametrized as a linear shift in $\epsilon$. The correct shift will be proportional to $\sqrt{\epsilon}$; however, this will not affect the bounds implied on the EFT coefficients. For a discussion of the entropy shift of very-near extremal black holes, see appendix \ref{app:vnearext}.

In order to avoid the divergence, we can instead consider a leading order solution taken slightly away from extremality.  As long as we are sufficiently close to extremality, the first term in (\ref{eq:Sshift}) will dominate the entropy shift.  Noting further that, at extremality, the numerator of (\ref{eq:drhde}) becomes proportional to the mass shift (\ref{mass_shift}) at fixed charge, we can rewrite (\ref{eq:Sshift}) as
\begin{align}
    &\left(\fft{\partial S}{\partial \epsilon}\right)_{Q,M}\!\!=\nonumber\\
    &-\fft{A}4\left(\fft{d-1}{(d-2)(M-(M_{\text{ext}})_0)}\left(\fft{\partial M}{\partial \epsilon}\right)_{Q,T=0}\!\!+\fft{d-1}{d-2}\rho+4c_1l^2R_{trtr}+2c_2l^2F_{tr}F_{tr}\right).
    \label{eq:ent_away_ext}
\end{align}
The deviation away from extremality can be written in terms of the leading order temperature,
\begin{align}
    4 \pi T_0 =|g'((r_h)_0)|_{\epsilon = 0} = \fft{(d-2)(M-(M_{\text{ext}})_0)}{V(d-1)(r_h)_0^{d-1}}\, .
\end{align}
The total shift to the entropy is then given by 
\begin{align}
    \left(\fft{\partial S}{\partial\epsilon}\right)_{Q,M}=-\fft1{T_0}\left(\fft{\partial M}{\partial \epsilon}\right)_{Q,T=0}-\fft{A}4\left(\fft{d-1}{d-2}\rho+4c_1l^2R_{trtr}+2c_2l^2F_{tr}F_{tr}\right)\,.
\end{align}
Finally, as $T_0 \rightarrow 0$ we reproduce the relation
\cite{Cheung:2018cwt, Goon:2019faz}
\begin{align}
    \left(  \frac{ \partial M}{\partial \epsilon} \right)_{Q, T = 0} =- T_0 \left(  \frac{ \partial S }{\partial \epsilon} \right)_{Q, M} \, .
\label{eq:GPshift}
\end{align}
Note that this relation was obtained using only the general feature that the corrected geometry may be written in terms of a shift $\Delta g$ to the radial function $g(r)$.  In particular, we never had to use the explicit form of $\Delta g$ given in (\ref{eq:Deltag}).

\subsection{Explicit Results for the Entropy Shifts}

\noindent In order to compare with the next section, we include some explicit results for the mass shifts. In section \ref{sec:posent}, we will see what constraints may be placed on the EFT coefficients by imposing that entropy shift is positive. We'll use the mass shift here, to remove the factor of $T_0$. The entropy shift is positive when the mass shift at constant charge is negative. It is easy to see that the shifts here are positive when all the coefficients are positive. 

\noindent For AdS${}_4$, we find:
\begin{align}
\label{TdS4}
    \begin{split}
            T_0 \Delta S = \frac{1}{5 r_h l^2}\Big( 4 c_1 (l^2 + 3 r_h^2)^2  + 2 c_2  (l^2 + 3 r_h^2) (l^2 + 18 r_h^2) + 8 (2 c_3 + c_4 ) (l^2 + 3 r_h^2)^2  \Big) \, .
    \end{split}
\end{align}
For AdS${}_5$, we get:
\begin{align}
\label{TdS5}
    \begin{split}
            T_0 \Delta S =  \frac{\pi}{16l^2}  \Big( & c_1 (31 l^4 + 128 l^2 r_h^2 + 138 r_h^4)\\
    & \qquad + c_2 24 (l^2 + 2 r_h^2) (l^2 + 6 r_h^2)+ (2 c_3 + c_4 ) 72 (l^2 + 2 r_h^2)^2 \Big) \, .
    \end{split}
\end{align}
AdS${}_6$:
\begin{align}
    \begin{split}
            T_0 \Delta S = \frac{2 \pi}{99 l^2}  \Big( & c_1 r_h (369 l^4 + 1263 l^2 r_h^2 + 1124 r_h^4)\\
    & \qquad  + c_2 4 r_h (3 l^2 + 5 r_h^2) (27 l^2 + 100 r_h^2)+ (2 c_3 + c4 ) 96 r_h (3 l^2 + 5 r_h^2)^2 \Big) \, .
    \end{split}
\end{align}
AdS${}_7$:
\begin{align}
    \begin{split}
            T_0 \Delta S =  \frac{\pi^2}{224 l^2}  \Big( & c_1 \, (1384 l^4 r_h^2 + 4236 l^2 r_h^4 + 3345 r_h^6) \\
    & \qquad  + c_2 \, 40 (2 l^2 + 3 r_h^2) (16 l^2 + 45 r_h^2) + (2 c_3 + c4 ) \, 800 (2l^2 + 3 r_h^2)^2 \Big) \, .
    \end{split}
\end{align}

\section{Thermodynamics from the On-Shell Euclidean Action}
\label{sec:euc}

\noindent The ultimate goal of this paper is to determine the leading higher-derivative corrections to relations between certain global properties of black hole solutions. These relations are of a thermodynamic nature, and arise by taking various derivatives of the free-energy corresponding to the appropriate ensemble. As is well-known \cite{Hawking:1979ig}, the classical free-energy of a black hole can be calculated using the saddle-point approximation of the Euclidean path integral with appropriate boundary conditions. In the \textit{Gibbs} or \textit{grand canonical ensemble}, the appropriate quantity is the Gibbs free-energy, which may be calculated from the on-shell Euclidean action
\begin{equation}
  \beta G(T,\Phi) = I_E[g^E_{\mu\nu}\left(T,\Phi\right),A^E_\mu(T,\Phi)],
\end{equation}
where $\beta = T^{-1}$, and  $g^E_{\mu\nu}\left(T,\Phi\right)$ and $A^E_\mu(T,\Phi)$ are Euclideanized solutions to the classical equations of motion with temperature $T$ and potential $\Phi$. Similarly in the \textit{canonical ensemble} the corresponding quantity is the Helmholtz free-energy, given by
\begin{equation}
  \beta F(T,Q) = I_E[g^E_{\mu\nu}\left(T,Q\right),A^E_\mu(T,Q)],
\end{equation}
where $g^E_{\mu\nu}\left(T,Q\right)$ and $A^E_\mu(T,Q)$ are Euclideanized solutions with temperature $T$ and electric charge $Q$. In both expressions, $I_E$ is the \textit{renormalized} Euclidean on-shell action.

The Euclidean action with cosmological constant is IR divergent when evaluated on a solution.  However, it may be given a satisfactory finite definition by first regularizing the integral with a radial cutoff $R$. To render the variation principle well-defined on a spacetime with boundary we must add an appropriate set of Gibbons-Hawking-York (GHY) \cite{York:1972sj,Gibbons:1976ue} (in the case of the canonical ensemble, also Hawking-Ross \cite{Hawking:1995ap}) terms \textit{in addition to} a set of boundary counterterms. The complete on-shell action then consists of three contributions
\begin{equation}
    I_{E} = I_{\text{bulk}} + I_{\text{GHY}} + I_{\text{CT}}.
\end{equation}
If the counterterms are chosen correctly, they will cancel the divergence of the bulk and Gibbons-Hawking-York terms, rendering the results finite as $R\rightarrow \infty$. In AdS there is a systematic approach to generating such counterterms via the method of \textit{holographic renormalization} \cite{Henningson:1998gx,Balasubramanian:1999re,Emparan:1999pm}; since the logic of this approach is well-described in detail elsewhere (see e.g.~\cite{Skenderis:2002wp}) we will not review it further, but simply make use of known results. Explicit expressions for the needed GHY and counterterms (including the four-derivative corrections used in this paper) valid in $\text{AdS}_d$, $d=4,5,6$ can be found in \cite{Liu:2008zf,Cremonini:2009ih}.

Once the free-energy is calculated, the remaining thermodynamic quantities can be determined straightforwardly by using the definitions of the free-energies and the first-law of black hole thermodynamics
\begin{equation}
  F = E-TS,\hspace{5mm} G= E-TS-\Phi Q, \hspace{5mm} dE = TdS+\Phi dQ.
\end{equation}
The expressions calculated using these Euclidean methods should agree with the Lorentzian or geometric calculations in the previous section.  Note, however, that there is a bit of a subtlety with the notion of black hole mass here, as the thermodynamic relations are for the energy $E$ of the system. In holographic renormalization, there is always an ambiguity in the addition of finite counterterms that shift the value of the on-shell action. The standard approach is to fix the ambiguity by demanding that even-dimensional global AdS has zero vacuum energy while odd-dimensional global AdS has non-zero vacuum energy that is interpreted as a Casimir energy in the dual field theory.  In this case the thermodynamic energy is the sum of the black hole mass and the Casimir energy
\begin{equation}
    E=M+E_c,
\end{equation}
and the mass $M$ of the black hole is only obtained after subtracting out the Casimir energy contribution, as we did in section \ref{sec:geo}.

The purpose of introducing this alternative approach is not just to give a cross-check on the results of the previous section, but also to verify a recent general claim by Reall and Santos \cite{Reall:2019sah}. In this paper, the $\mathcal{O}(\epsilon)$ corrections we are considering can be calculated by first evaluating the free-energy or on-shell action at the same order. Naively, this would require evaluating three contributions
\begin{align}
  I_E[g^E_{\mu\nu},A^E_\mu] &= I^{(2)}_E[g^{(2)E}_{\mu\nu},A^{(2)E}_\mu] + \epsilon \left(\frac{\partial}{\partial \epsilon} I^{(2)}_E[g^{(2)E}_{\mu\nu}+\epsilon g^{(4)E}_{\mu\nu} ,A^{(2)E}_\mu+\epsilon A^{(4)E}_\mu] \right)\biggr\vert_{\epsilon=0} \nonumber\\
                            &\hspace{5mm} +\epsilon I^{(4)}_E[g^{(2)E}_{\mu\nu},A^{(2)E}_\mu] + \mathcal{O}\left(\epsilon^2\right),
\end{align}
where $(2)$ and $(4)$ denote two and four derivative terms in the action and their corresponding perturbative contributions to the solution.  The central claim in \cite{Reall:2019sah} is that the first term at $\mathcal{O}(\epsilon)$ is actually \textit{zero}, and that therefore we do not need to explicitly calculate the $\mathcal{O}(\epsilon)$ corrections to the equations of motion. For black hole solutions of the type considered in this paper, we can evaluate the leading corrections without much difficulty, but for more general situations with less symmetry this may not be possible. In such a case the Euclidean method is more powerful, as has recently been demonstrated with calculation of corrections involving angular momentum \cite{Cheung:2019cwi} or dilaton couplings \cite{Loges:2019jzs}. 

Although the result of \cite{Reall:2019sah} was demonstrated in the grand canonical ensemble, it is straightforward to see that it implies an identical claim about the leading corrections in the canonical ensemble. While the quantities of interest can be extracted from either, the explicit expressions encountered in the latter are usually far simpler and therefore more convenient. Recall that we can change ensemble by a Legendre transform of the free-energy
\begin{equation}
  F(T,Q) = G(T,\Phi(Q)) + \Phi(Q) Q, \hspace{5mm} Q = -\left(\frac{\partial G}{\partial \Phi}\right)_{T},
\end{equation}
where the right-hand-side is defined in terms of the implicit inverse function $\Phi(Q)$. At fixed $T$ and $Q$, the potential $\Phi$ receives corrections from the higher-derivative interactions, and so, expanding the right-hand-side to $\mathcal{O}(\epsilon)$, we have
\begin{align}
  F(T,Q) &= G^{(2)}(T,\Phi^{(2)}(Q)) + \epsilon\left(\frac{\partial}{\partial \epsilon}G^{(2)}(T,\Phi^{(2)}(Q)+\epsilon\Phi^{(4)}(Q))\right)\biggr\vert_{\epsilon=0} \nonumber\\
&\hspace{5mm}+ \epsilon G^{(4)}(T,\Phi^{(2)}(Q)) + \Phi^{(2)}(Q)Q +\epsilon \Phi^{(4)}(Q)Q +\mathcal{O}\left(\epsilon^2\right).
\end{align} 
Recognizing that 
\begin{equation}
  \left(\frac{\partial}{\partial \epsilon}G^{(2)}(T,\Phi^{(2)}(Q)+\epsilon\Phi^{(4)}(Q))\right)\biggr\vert_{\epsilon=0} = \Phi^{(4)}(Q)\left(\frac{\partial G^{(2)}}{\partial \Phi}\right)_T\biggr\vert_{\Phi=\Phi^{(2)}(Q)} = -\Phi^{(4)}(Q)Q,
\end{equation}
we see that the leading correction to the Helmholtz free energy is simply given by 
\begin{equation}
   F(T, Q) = F^{(2)}(T, Q) +\epsilon G^{(4)}(T,\Phi^{(2)}(Q)) + \mathcal{O}\left(\epsilon^2\right).
\end{equation}
In terms of the on-shell Euclidean action, using the result of Reall and Santos, this is then equivalent to
\begin{equation}
\label{F4}
  F^{(4)}(T,Q) = \frac{1}{\beta}I^{(4)}_E\left(g^{(2)E}_{\mu\nu}\left(T,Q\right),A^{(2)E}_\mu(T,Q)\right),
\end{equation}
where here $I_E^{(4)}$ denotes the contribution of the four-derivative terms to the renormalized on-shell action. Note that this includes potential four-derivative Gibbons-Hawking-York terms, but as this argument makes clear, will not include any additional Hawking-Ross terms. This expression is the analogue of the Reall-Santos result, but in the canonical ensemble. It says that the leading correction to the Helmholtz free-energy is given by evaluating the four-derivative part of the renormalized on-shell action on a solution to the two-derivative equations of motion with temperature $T$ and charge $Q$. 

Below we will give a brief review of the well-known thermodynamic relations at two-derivative order, and then using the above result we will calculate the leading corrections and verify explicitly that they agree with the results of the previous section.

\subsection{Two-Derivative Thermodynamics}

\noindent As described above, the regularized on-shell action has a bulk as well as various boundary contributions. At two-derivative order and in $d$-dimensions these have the explicit form
\begin{align}
  I^{(2)}_{\text{bulk}} &= -\frac{1}{16 \pi } \int d^{d+1} x \sqrt{g} \Big(  \frac{d (d-1)}{l^2} + R -\frac{1}{4} F^2 \Big), \nonumber\\
  I^{(2)}_{\text{GHY}} &= -\frac{1}{8 \pi } \int d^{d} x \sqrt{h} K, \nonumber\\
  I^{(2)}_{\text{CT}} &= \frac{1}{8 \pi } \int d^{d} x \sqrt{h} \left( \frac{d - 1}{l} + \frac{l}{2 (d - 2)} \mathcal{R}\right),
\end{align}
where $h_{ab}$ and $\mathcal{R}_{ab}$ are the metric and Ricci tensor of the induced geometry on the boundary at $r = R$. Note that in $I^{(2)}_{\text{CT}}$ we have included the minimal set of counterterms necessary to cancel the IR divergence in $d=3$ and $d=4$. For $d>4$, additional counterterms beginning at quadratic order in the boundary Riemann tensor are necessary to cancel further divergences. 

The regularized bulk action has a well-defined variational principle provided that $\delta A_{a} = 0$ at $r = R$. This amounts to holding $\Phi$ fixed, and thus it corresponds to boundary conditions compatible with the grand canonical ensemble. For many applications, we will want to hold the charge fixed. From a thermodynamic point of view, we want to use the extensive quantity $Q$ instead of the intensive $\Phi$, so we must compute the Helmholtz free energy instead of the Gibbs free energy. Holding $Q$ fixed requires different boundary conditions, and in particular
the further addition of a Hawking-Ross boundary term \cite{Hawking:1995ap} 
\begin{align}
    I^{(2)}_{\text{HR}} =  \frac{1}{16 \pi} \int d^{d} x \sqrt{h} n_\mu F^{\mu b} A_b \, ,
\end{align}
where $n_\mu$ is the normal vector on the boundary and $A_a$ is the pull-back of the gauge potential. To summarize, the total two-derivative on-shell action
\begin{equation}
  I^{(2)}_E = I^{(2)}_{\text{bulk}}+I^{(2)}_{\text{GHY}}+I^{(2)}_{\text{HR}}+I^{(2)}_{\text{CT}},
\end{equation}
evaluated on the Euclideanized solution to the two-derivative equations of motion
\begin{align}
    \begin{split}
        ds_E^2 =& f(r) d\tau^2 + g(r)^{-1} dr^2 + r^2 d \Omega_{d-1}^2 \, , \qquad f(r) = g(r) = 1 - \frac{m}{r^{d-2}} + \frac{q^2}{4 r^{2d-4}} + \frac{r^2}{l^2}, \\
        & \quad A_E = i\left( - \frac{1}{c} \frac{q}{r^{d-2}} + \Phi \right) d\tau,  \qquad c = \sqrt{ \frac{ 2(d-2)}{(d-1)} }, \qquad \Phi = \frac{1}{c} \frac{q}{l^{d-2}\nu^{d-2}}\, ,
    \end{split}
\end{align}
is equal to $\beta F^{(2)}(T,Q)$, where $F^{(2)}$ is the two-derivative contribution to the Helmholtz free-energy. In the above we have introduced the dimensionless variable $\nu\equiv (r_h)_0/l$, where $(r_h)_0$ is the location of the outer-horizon of the two-derivative solution with temperature $T$ and charge $Q$. Note also that here, and for the remainder of this section, we will consider only spherical $k=1$ black holes. Since $\nu$ satisfies $f(\nu)=0$, we can solve for the parameter $m$ as
\begin{equation}
  m = \nu^{d-2}+\frac{q^2}{4\nu^{d-2}}+\frac{\nu^d}{l^2}.
\end{equation}
In the Euclidean approach to calculating the leading corrections to the thermodynamics, it will prove natural to continue to use $\nu$ and $q$ to parametrize the space of black hole solutions, even when the four-derivative corrections are included. This means that it is also natural to write all thermodynamic quantities in these variables, which requires the use of standard thermodynamic derivative identities to rewrite derivatives. Recall that the parameter $q$ and the physical charge $Q$ are not the same, but are related by an overall constant given in (\ref{eq:Qdef}).  Therefore holding $Q$ fixed is the same as holding $q$ fixed. Explicitly, the two-derivative free-energy calculated in this way in $\text{AdS}_{4}$ is given by
\begin{equation}
\label{F2d3}
 F_{d=3}^{(2)}(q,\nu) = -\frac{l \nu ^3}{4}+\frac{l \nu }{4}+\frac{3 q^2}{16 l \nu } ,
\end{equation}
and in $\text{AdS}_5$ by
\begin{equation}
\label{F2d4}
 F_{d=4}^{(2)}(q,\nu) = -\frac{1}{8} \pi  l^2 \nu ^4+\frac{1}{8} \pi  l^2 \nu ^2+\frac{5 \pi  q^2}{32 l^2 \nu
   ^2}+\frac{3 \pi  l^2}{32} .
\end{equation}
Once the free-energy is calculated, the entropy and energy are given by
\begin{align}
S = - \left( \frac{\partial F}{ \partial T} \right)_Q, \hspace{5mm} E = F+TS.
\end{align}
In terms of our natural variables, we can re-express the entropy as
\begin{equation}
  S(q,\nu) = \left(\frac{\partial F}{\partial \nu}\right)_q \left[\left(\frac{\partial T}{\partial \nu}\right)_q\right]^{-1},
\end{equation}
where the temperature is given by
\begin{equation}
  T(q,\nu) = \frac{(d-2)q ^2 l^{1-d}\nu^{1-d}}{4\pi}+ \frac{(d-1)\nu^2+d-2}{4\pi l}.
\end{equation}
Note that this expression is exact, meaning it does not receive corrections when we include the four-derivative interactions. It is therefore useful to introduce the function
\begin{equation}
\label{qext}
    q^2_{\text{ext}}(\nu) = -\frac{2 \left(d \nu ^2+d-\nu ^2-2\right) (l \nu )^{d-2}}{(d-2) },
\end{equation}
such that taking the limit $q^2\rightarrow  q^2_{\text{ext}}(\nu)$ is equivalent to taking the extremal limit $T\rightarrow 0$. 

If we extract the energy $E=F+TS$ from the expressions (\ref{F2d3}) and (\ref{F2d4}), we find that it agrees with the mass, (\ref{eq:massdef}), for $\text{AdS}_4$ but not $\text{AdS}_5$. This is not surprising as the thermodynamic energy $E$ and mass $M$ of the black hole in AdS$_5$ differ by a Casimir energy contribution that is independent of $q$ and $\nu$.  We can, of course remove the Casimir energy by the addition of \textit{finite boundary counterterms}, or equivalently by a change in holographic renormalization scheme. The expression (\ref{F2d4}) is calculated in a \textit{minimal subtraction} scheme, in which the possible finite counterterms are zero and the Casimir energy is present.

Physically, it is useful work in a scheme in which the energy $E$ coincides with the mass $M$ of the black hole, without a Casimir contribution. In such a \textit{zero Casimir} scheme, the energy of pure $\text{AdS}_5$ is defined to be zero. Calculating the free-energy from the on-shell action of pure $\text{AdS}_5$ with generically parametrized four-derivative counterterms we find that this scheme requires the following modification from the minimal subtraction counterterms 
\begin{equation}
    I_{\text{CT}}^{(2)} \longrightarrow I_{\text{CT}}^{(2)}+\frac{1}{8 \pi } \int d^{4} x \sqrt{h} \left( -\frac{l^3 }{96}\right) \mathcal{R}^2.
\end{equation}
The free energy calculated with this modified on-shell action agrees exactly with the expectation using (\ref{eq:massdef}). Note that the entropy, since it is given by a derivative of the free-energy, is independent of the choice of scheme. The zero Casimir scheme is a physically motivated choice, but certainly not unique. 

\subsection{Four-Derivative Corrections to Thermodynamics}

\noindent To evaluate the four-derivative corrections we make use of the result (\ref{F4}). As in the two-derivative contribution, the on-shell action is properly defined by a regularization and renormalization procedure. For the operators in (\ref{Action}) with Wilson coefficients $c_2$, $c_3$ and $c_4$ the required $I_{\text{bulk}}^{(4)}$ contribution is actually finite, while for the term in (\ref{Action}) proportional to $c_1$, we must again regularize and renormalize by adding infinite boundary counterterms. The required explicit expressions, as well as the complete set of four-derivative GHY terms, can be found in \cite{Liu:2008zf,Cremonini:2009ih}. 
The calculation is otherwise identical to the two-derivative contribution described above, and in $\text{AdS}_4$ we find
\begin{align}
    F^{(4)}_{d=3}(q,\nu) = &c_1 \left(-\frac{ \left(20 l^4 \nu ^4-5 l^2 \nu ^2 q^2+q^4\right)}{20 l^5 \nu
   ^5}-\frac{3 \nu   }{l}\right)+\frac{c_2 q^2  \left(l^2 \left(20 l^2
   \nu ^2-7 q^2\right)-60 l^4 \nu ^4\right)}{80 l^7 \nu ^5}\nonumber\\
    &    -\frac{c_3 q^4 }{5
   l^5 \nu ^5}-\frac{c_4 q^4 }{10 l^5 \nu ^5}.
\end{align}
The complete free-energy, up to $\mathcal{O}(\epsilon^2)$ contributions, is then given by
\begin{equation}
    F_{d=3}(q,\nu) = F_{d=3}^{(2)}(q,\nu)+\epsilon  F_{d=3}^{(4)}(q,\nu) + \mathcal{O}(\epsilon^2).
\end{equation}
From this explicit expression we can then calculate the entropy
\begin{align}
    S_{d=3}&=\pi  l^2 \nu ^2-\frac{4 \pi  c_1 \epsilon  \left(4 l^4 \nu ^4 \left(1-3 \nu ^2\right)-3 l^2 \nu ^2
   q^2+q^4\right)}{4 l^2 \left(3 \nu ^2-1\right) \nu ^4+3 \nu ^2 q^2}-\frac{\pi  c_2 q^2
   \epsilon  \left(12 l^2 \nu ^2 \left(\nu ^2-1\right)+7 q^2\right)}{4 l^2 \left(3 \nu
   ^2-1\right) \nu ^4+3 \nu ^2 q^2}\nonumber\\
   & \hspace{5mm}   -\frac{16 \pi  c_3 q^4 \epsilon }{4 l^2 \left(3 \nu
   ^2-1\right) \nu ^4+3 \nu ^2 q^2}-\frac{8 \pi  c_4 q^4 \epsilon }{4 l^2 \left(3 \nu
   ^2-1\right) \nu ^4+3 \nu ^2 q^2}+\mathcal{O}(\epsilon^2),
\end{align}
and mass (which coincides with the thermal energy)
\begin{align}
   M_{d=3} &=\frac{1}{2} l \left(\nu ^3+\nu \right)+\frac{q^2}{8 l
   \nu }+\frac{c_1 q^4 \epsilon  \left(q^2-4 l^2 \nu ^2 \left(9 \nu ^2+2\right)\right)}{40 l^5
   \left(3 \nu ^2-1\right) \nu ^7+30 l^3 \nu ^5 q^2}\nonumber\\
   &\hspace{5mm}+\frac{c_2 q^2 \epsilon  \left(80 l^4
   \nu ^4 \left(-9 \nu ^4+6 \nu ^2+1\right)-8 l^2 \nu ^2 \left(39 \nu ^2+7\right) q^2+7
   q^4\right)}{40 l^3 \nu ^5 \left(4 l^2 \nu ^2 \left(3 \nu ^2-1\right)+3 q^2\right)}\nonumber\\
   &\hspace{5mm}+\frac{2
   c_3 q^4 \epsilon  \left(q^2-4 l^2 \nu ^2 \left(9 \nu ^2+2\right)\right)}{5 l^3 \nu ^5
   \left(4 l^2 \nu ^2 \left(3 \nu ^2-1\right)+3 q^2\right)}+\frac{c_4 q^4 \epsilon 
   \left(q^2-4 l^2 \nu ^2 \left(9 \nu ^2+2\right)\right)}{5 l^3 \nu ^5 \left(4 l^2 \nu ^2
   \left(3 \nu ^2-1\right)+3 q^2\right)}+\mathcal{O}(\epsilon^2).
\end{align}
Taking the extremal limit we find the following expression for the mass shift
\begin{align}
    \label{massshift}
    (\Delta M_{d=3})_{Q,T=0} = &-\frac{4 c_1 l \left(3 \nu ^2+1\right)^2}{5 \nu }-\frac{2 c_2 l (3 \nu ^2+1)(18 \nu ^2+1)}{5 \nu }\nonumber\\
    &   -\frac{16 c_3 l \left(3 \nu ^2+1\right)^2}{5 \nu }-\frac{8
   c_4 l \left(3 \nu ^2+1\right)^2}{5 \nu },
\end{align}
which agrees exactly with (\ref{TdS4}). Strictly, the two expressions are parameterized in terms of different variables ($\nu$ the uncorrected horizon vs. $r_h$ the corrected horizon), but these differ by $\mathcal{O}(\epsilon)$, and so when we take $\epsilon \rightarrow 0$ the two functions are the same.

Similarly we can calculate the shift in the microcanonical entropy, which will be important in the subsequent section for analyzing conjectured bounds on the Wilson coefficients. The actual expression is given in (\ref{entshiftads4}), and can be calculated straightforwardly using standard thermodynamic derivative identities 
\begin{equation}
\label{microSformula}
    (\Delta S)_{Q,E} = \lim_{\epsilon\rightarrow 0}\left[\left(\frac{\partial S}{\partial \epsilon}\right)_{q,\nu}- \left(\frac{\partial E}{\partial \epsilon}\right)_{q,\nu}\frac{\left(\frac{\partial S}{\partial \nu}\right)_{q}}{\left(\frac{\partial E}{\partial \nu}\right)_{q}}\right].
\end{equation}
The calculation for $\text{AdS}_5$ is similar, but in this case we have to be cautious about the Casimir energy. We calculate the free-energy in the physically motivated zero Casimir scheme.  To do so, we again fix the finite counterterms by evaluating the four-derivative on-shell action on pure $\text{AdS}_5$. Requiring the Casimir energy to vanish requires the following modification from the minimal subtraction counterterm action  
\begin{equation}
    I_{\text{CT}}^{(4)} \longrightarrow I_{\text{CT}}^{(4)}+\frac{1}{8 \pi } \int d^{4} x \sqrt{h} \left(-\frac{5 c_1 l^3}{48}\right) \mathcal{R}^2.
\end{equation}
Using this we calculate the four-derivative contribution to the renormalized free-energy 
\begin{align}
    F_{d=4}^{(4)} = &\frac{1}{256} \pi  c_1 \left(-\frac{43 q^4}{l^8 \nu ^8}+\frac{24 \left(5 \nu
   ^2+8\right) q^2}{l^4 \nu ^4}-32 \left(13 \nu ^4+41 \nu ^2+18\right)\right)\nonumber\\
   &    +\frac{3 \pi 
   c_2 \left(8 l^4 \nu ^4 q^2-3 q^4\right)}{32 l^8 \nu ^8}-\frac{9 \pi  c_3
   q^4}{16 l^8 \nu ^8}-\frac{9 \pi  c_4 q^4}{32 l^8 \nu ^8}.
\end{align}
We also obtain the entropy
\begin{align}
    S_{d=4} =&\frac{1}{2} \pi ^2 l^3 \nu ^3+\frac{\pi ^2 c_1 \epsilon  \left(8 l^8 \left(26 \nu ^2+41\right) \nu ^{10}+6 l^4
   \left(5 \nu ^2+16\right) \nu ^4 q^2-43 q^4\right)}{4 l^3 \nu ^3 \left(4 l^4 \left(2 \nu
   ^2-1\right) \nu ^4+5 q^2\right)}\nonumber\\
   &+\frac{6 \pi ^2 c_2 \epsilon  \left(4 l^4 \nu ^4
   q^2-3 q^4\right)}{l^7 \left(8 \nu ^9-4 \nu ^7\right)+5 l^3 \nu ^3 q^2}-\frac{36 \pi ^2
   c_3 q^4 \epsilon }{l^7 \left(8 \nu ^9-4 \nu ^7\right)+5 l^3 \nu ^3 q^2}\nonumber\\
   &    -\frac{18
   \pi ^2 c_4 q^4 \epsilon }{l^7 \left(8 \nu ^9-4 \nu ^7\right)+5 l^3 \nu ^3
   q^2}+\mathcal{O}\left(\epsilon^2\right),
\end{align}
and mass
\begin{align}
    M_{d=4} =&\frac{3 \pi  \left(4 l^4 \left(\nu ^2+1\right) \nu ^4+q^2\right)}{32 l^2
   \nu ^2}\nonumber\\
   &+ c_1\left[\frac{\pi  \epsilon  \left(384 l^{12} \left(\nu ^2+1\right) \left(26 \nu ^4+23
   \nu ^2+6\right) \nu ^{12}-32 l^8 \left(27 \nu ^4+32 \nu ^2+18\right) \nu ^8 q^2\right)}{256 l^8 \nu ^8 \left(4 l^4
   \left(2 \nu ^2-1\right) \nu ^4+5 q^2\right)}\right.\nonumber\\
   &\hspace{10mm}+\left.\frac{\pi  \epsilon  \left(-4 l^4
   \left(684 \nu ^2+253\right) \nu ^4 q^4+129 q^6\right)}{256 l^8 \nu ^8 \left(4 l^4
   \left(2 \nu ^2-1\right) \nu ^4+5 q^2\right)}\right]\nonumber\\
   &+\frac{3 \pi  c_2 q^2 \epsilon 
   \left(32 l^8 \left(10 \nu ^2+3\right) \nu ^8-4 l^4 \left(54 \nu ^2+19\right) \nu ^4
   q^2+9 q^4\right)}{32 l^8 \nu ^8 \left(4 l^4 \left(2 \nu ^2-1\right) \nu ^4+5
   q^2\right)}\nonumber\\
   &+\frac{9 \pi  c_3 q^4 \epsilon  \left(3 q^2-4 l^4 \nu ^4 \left(18 \nu
   ^2+7\right)\right)}{16 l^8 \nu ^8 \left(4 l^4 \left(2 \nu ^2-1\right) \nu ^4+5
   q^2\right)}\nonumber\\
   &+\frac{9 \pi  c_4 q^4 \epsilon  \left(3 q^2-4 l^4 \nu ^4 \left(18 \nu
   ^2+7\right)\right)}{32 l^8 \nu ^8 \left(4 l^4 \left(2 \nu ^2-1\right) \nu ^4+5
   q^2\right)} +\mathcal{O}\left(\epsilon^2\right).
\end{align}
The extremal mass shift is given by
\begin{align}
    (\Delta M_{d=4})_{Q,T=0} = &-\frac{1}{16} \pi  c_1 \left(138 \nu ^4+128 \nu ^2+31\right)-\frac{3}{2} \pi 
   c_2  \left(2 \nu ^2+1\right) \left(6 \nu ^2+1\right)\nonumber\\
   &    -9 \pi  c_3 \left(2 \nu
   ^2+1\right)^2-\frac{9}{2} \pi  c_4 \left(2 \nu ^2+1\right)^2 \, ,
\end{align}
which agrees exactly with the result (\ref{TdS5}). Likewise we can calculate the correction to the microcanonical entropy using (\ref{microSformula}), the explicit expression is given in (\ref{dSde5}).

\section{Constraints from Positivity of the Entropy Shift}
\label{sec:posent}

\noindent Having derived the general entropy shift at fixed mass, we may now determine what constraints on the EFT coefficients are implied by the assumption that it is positive. 
Recall that the argument of \cite{Cheung:2018cwt}
for the positivity of the entropy
shift 
assumes the existence of a number of quantum fields $\phi$ with mass $m_{\phi}$, heavy enough so that they can be safely integrated out. 
In particular, such fields are assumed to couple to the graviton and photon in such a way that, after 
being integrated out, they generate \textit{at tree-level} the higher-dimension operators we are considering (with the corresponding operator coefficients scaling as $c_i \sim 1/m_{\phi}$). This assumption is essential to the proof; it may be that the entropy shift is universally positive (see \cite{Cheung:2019cwi} for a number of examples), but proving such a statement for non-tree-level completions would require a different argument from the one laid out here.

We revisit the logic of \cite{Cheung:2018cwt} in the context of flat space, before discussing how it may be extended to AdS asymptotics, and denote the Euclidean on-shell action of the theory that includes the heavy scalars $\phi$  by $I_{\text{UV}}[g, A, \phi]$.
First, note that when the scalars are set to zero and are non-dynamical, the action reduces to that of the pure Einstein-Maxwell theory, 
\begin{align}
    I_{\text{UV}}[g, A, 0] = I^{(2)}[g, A] \, .
\end{align}
This is a statement relating the value of the functionals $I_{\text{UV}}$ and $I^{(2)}$ (the two-derivative action) when we pick particular configurations for the fields. These fields may or may not be solutions to the equations of motion. 
Next, consider the corrected action, $I_C = I^{(2)} + I^{(4)}$, and note that it obeys  
\begin{align}
    I_C[g + \Delta g, A + \Delta A] \simeq I_{\text{UV}}[g, A, \phi] \, .
\end{align}
Here we have in mind that the fields are valid solutions of the respective theories, \emph{i.e.} 
$[g, A, \phi]$ is a solution of the $\text{UV}$ theory and  $[g + \Delta g, A + \Delta A]$ is a solution to the four-derivative corrected theory. The $\text{UV}$
theory and that with an infinite series of higher-derivative corrections should have exactly the same partition function; therefore, this expression is an equality up to quantum corrections and corrections that are $\mathcal{O}(\epsilon^2)$. 
Finally, let us choose $[g, A, \phi]$ to be solutions of the $\text{UV}$ theory with charge $Q$ and temperature $T$, and 
$[g_0, A_0]$ to be field configurations in the pure Einstein-Maxwell theory \emph{with the same charge and temperature} as those of the UV theory. One then finds the following inequality,
\begin{align}
    I_C[g + \Delta g, A + \Delta A]_{T, Q} \simeq I_{\text{UV}}[g, A, \phi]_{T, Q} < I_{\text{UV}}[g_0, A_0, 0]_{T, Q} = I^{(2)}[g_0, A_0]_{T, Q} \, .
    \label{action_ineqality}
\end{align}
Since $[g, A, \phi]$ is a solution of the UV theory,  it extremizes the action. To ensure the inequality that appears in (\ref{action_ineqality}), one must further require that this solution is a \textit{minimum} of the action. The inequality then follows because $[g_0, A_0,0]$ is \emph{not} a solution to the equations of motion, for the same charge and temperature.
Finally, as long as one works in the same ensemble, the boundary terms will be the same for both actions and thus don't affect the argument.

In general, different theories will have different relationships between mass, charge, and temperature. We are interested in the entropy shift at fixed mass and charge. Therefore we must compare the two action functionals at different temperatures. For simplicity, we use $T_4/T_2$ for the temperature that corresponds to mass $M$ and charge $Q$ for the theory with/without higher-derivative corrections, respectively. Then we have:
\begin{align}
    \begin{split}
        F_C(Q, T_4) & < F_2(Q, T_4) ,\\
        F_C(Q, T_4) & < F_2(Q, T_2) + (T_4 - T_2) \partial_T F_2(Q, T_2), \\
        F_C(Q, T_4) & < F_2(Q, T_2) - (T_4 - T_2) S_2, \\
        M - S_4 T_4 & < M - S_2 T_2 - (T_4 - T_2) S_2 ,\\
         - S_4 T_4 & <  - T_4  S_2, \\
         \Delta S & > 0,
    \end{split}
\end{align}
at fixed $M$ and $Q$ (and in the zero Casimir energy scheme).

Now that we have outlined the argument in flat space, we can ask whether it can be immediately extended to AdS. 
One subtle point in the derivation outlined above is that  the free-energy is only finite after the subtraction of the free-energy of a reference background. In the flat space context, the contributions of such terms to the two actions are identical because the asymptotic charges are the same. Thus, this issue does not affect the validity of the argument.

In AdS, the story is a little different-- the free-energy is computed using holographic renormalization. Different counterterms are required to render the two-derivative action $I^{(2)}$ and the corrected action $I_C$ finite. 
Moreover, $I_{UV}$ may also require a different set of counterterms involving contributions from the scalar, and unlike the bulk contribution, there is no reason to expect that their on-shell values are less than their off-shell values. 
This is a potential hole in the positivity argument in AdS. Apart from this issue, the rest of the argument can be immediately applied to AdS.

\subsection{Thermodynamic Stability }

\noindent As we've seen, the above proof requires that the uncorrected backgrounds are minima of the action. Thermodynamically, this amounts to the condition that the black holes are stable under thermal and electrical fluctuations. This translates to the following requirements on the free-energies,
\begin{align}
        \left( \frac{ \partial^2 F}{\partial T^2} \right)_Q \leq 0, \qquad  \left( \frac{ \partial^2 G}{\partial T^2} \right)_{\Phi} \leq 0, \qquad \epsilon_T = \left( \frac{ \partial^2 F}{\partial Q^2} \right)_T \geq 0 \, .
\end{align}
These conditions may be rewritten in terms of the specific heat and permittivity of the black hole,
which can be used to determine, respectively, the 
thermal stability and electrical stability of the black hole \cite{Chamblin:1999tk, Chamblin:1999hg}. We'll ignore the specific heat at constant $\Phi$ now, as we are interested in the stability in the canonical ensemble, and consider 
\begin{align}
    C_Q = T \left( \frac{ \partial S}{\partial T} \right)_Q \geq 0, \qquad \epsilon_T = \left( \frac{ \partial Q}{\partial \Phi} \right)_T \geq 0 \, .
\end{align}
Positivity of the specific heat is equivalent to the statement that larger black holes should heat up and radiate more, while smaller ones should become colder and radiate less.
When the quantity $\epsilon_T$ is negative the 
black hole is unstable to electrical fluctuations, meaning that when more charge is placed into it, its chemical potential \textit{decreases}. We expect that it should instead increase, to make it more difficult to move a charge from outside to inside the black hole -- thus making it harder to move away from equilibrium \cite{Chamblin:1999hg}. We may compute these quantities using the results of the previous section. For AdS${}_4$, we find
\begin{align}
    C_Q =\frac{2 \pi l^2 \nu^2 (1 + 3 \nu^2) (2 - \xi) \xi}{2 - 6 \xi + 3 \xi^2 + 3 \nu^2 (4 - 6 \xi + 3 \xi^2)}, \qquad \epsilon_T =  \frac{(\xi - 2 ) \xi + 3 \nu^2 (2 - 2 \xi + \xi^2)}{\nu l \left( 2 - 6 \xi + 3 \xi^2 + 3 \nu^2 (4 - 6 \xi + 3 \xi^2)\right)} \, ,
    \label{specific_heats}
\end{align}
where we recall that $\nu = r_h / l$ and $Q = (1 - \xi) Q_{\text{ext}}$. These results have been obtained previously \emph{e.g.} in \cite{Shen:2005nu}. We find that both of these coefficients are positive when either 
\begin{equation}
\label{firstcondition}
    \nu< \nu^* =  \frac{1}{\sqrt{3}} \, , \qquad
    \xi< \xi^* = 1 - \sqrt{\frac{1 - 3 \nu^2}{1 + 3 \nu^2}}\, , 
\end{equation}
holds, or when
\begin{equation}
  \nu > \nu^* =  \frac{1}{\sqrt{3}} \, , 
  \qquad 0<\xi <1    \, ,
  \label{secondcondition}
\end{equation}
is satisfied.

Thus, for small black holes stability requires that the extremality parameter be less than some function of the radius, $\xi < \xi^*$. 
In particular, extremal black holes, for which $\xi \rightarrow 0$, are stable while neutral black holes, which correspond to $\xi \rightarrow 1$, are not. 
The implication of (\ref{secondcondition}) is that above a certain radius ($r_h >  l / \sqrt{3}$) all black holes are thermodynamically stable. 
\begin{figure}
    \includegraphics[scale=0.59]{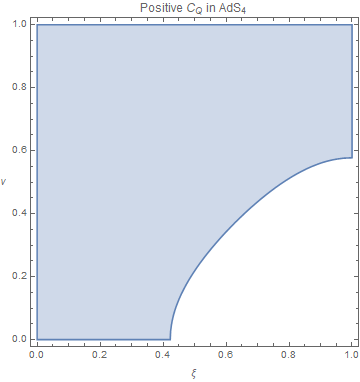}
    \includegraphics[scale=0.59]{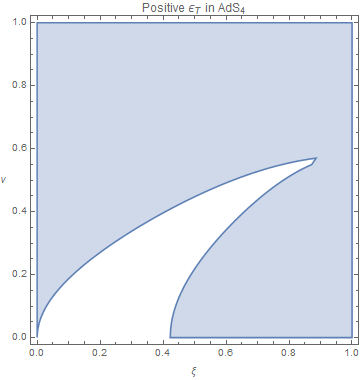}
    \caption{Blue represents the regions of parameter space where each quantity is positive.}
    \label{fig: stability} 
\end{figure}
This behavior is visible from Fig. \ref{fig: stability}, where we have plotted the allowed parameter space based on the $C_Q$ and $\epsilon_T$ conditions separately. This raises an interesting point in making contact with the flat space limit: if we require both parameters to be positive, there are no stable black holes at $\nu = 0$. Note that in \cite{Cheung:2018cwt} only $C_Q$ was considered. However, in applications involving AdS/CFT, we believe that both the specific heat and electrical permittivity should be taken into account.

Here we have only considered the leading-order stability. The higher-derivative corrections will shift the point where the specific heat crosses from positive to negative. However, in proving the extremality-entropy relation, we are only interested in the extremal limit, which is not affected by this consideration. In principal we could compute the order $\epsilon$ shifts to the stability conditions to obtain small corrections to the entropy bounds.

\subsection{Constraints on the EFT Coefficients}

\noindent The entropy shift in AdS${}_4$ for a black hole with an \emph{arbitrary size and charge} takes the following form,
\begin{align}
\begin{split}
        & \left(\frac{\partial S}{\partial \epsilon} \right)_{Q, M} \, = \, \frac{l(1 + 3 \nu^2)}{5 \nu T} \Big(c_1 \left(   4 - 6 \xi + 19 \xi^2 - 16 \xi^3 + 4 \xi^4 +12 \nu^2 (\xi-1)^4  \right) \\
        & \ \  + c_2 (\xi-1)^2 \left(   2 - 14 \xi + 7 \xi^2 + 3 \nu^2 (12 - 14 \xi + 7 \xi^2)  \right) +  8 (2 c_3 + c_4) (1 + 3 \nu^2) (\xi -1)^4 \Big) \, ,
        \label{entshiftads4}
\end{split}
\end{align}
where the temperature is given by the expression
$$T(r_h, \xi)  = - \frac{(1 + 3 \nu^2) (\xi-2) \xi}{4 \pi \nu l} \, .$$
We can see from the $\xi$ dependence of the latter that in the $\xi \rightarrow 0$ limit the shift to the entropy blows up. If we examine the leading part in $1 / \xi$, we find that it is proportional to the mass shifts we have computed above. Thus, in the extremal limit we have
\begin{align}
     \left(\frac{\partial S}{\partial \epsilon} \right)_{\xi \rightarrow 0} \, = \,  \frac{l^2}{5 r_h T} \Big( 4 c_1 (1 + 3 \nu^2)^2 + 2 c_2 (1 + 3 \nu^2) (1 + 18 \nu^2) + 8 (2 c_3 + c4)  (1 +  3 \nu^2)^2  \Big) \, .
\end{align}
It is also interesting to note that in the chargeless limit $\xi \rightarrow 1$ the dependence of (\ref{entshiftads4}) on $c_2, c_3$ and $c_4$ drops out entirely, and  we are left with an entropy shift 
of the simple form
\begin{align}
     \left(\frac{\partial S}{\partial \epsilon} \right)_{\xi \rightarrow 1} \, = \,  \frac{l}{\nu T} c_1 \left( 1 + 3 \nu^2 \right) \,.
\end{align}
Our results above show that large black holes are stable in the chargeless limit. Therefore, under the assumption that the four-derivative corrections yield a positive entropy shift for all possible stable black hole backgrounds, we find
\begin{align}
    c_1 \geq 0 \, .
\end{align}

\begin{figure}
    \includegraphics[scale=0.59]{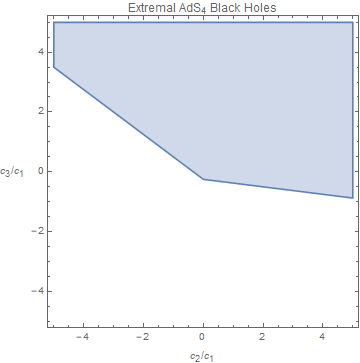}
    \includegraphics[scale=0.59]{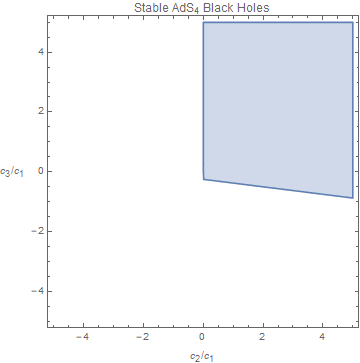}
    \caption{Blue regions are allowed after imposing that the entropy shift is positive. (Left): Allowed region after imposing that extremal black holes have positive entropy shift (Right):  Allowed region after imposing that all stable black holes have positive entropy shift}
    \label{AdS4_exclusion} 
\end{figure}

In Fig. \ref{AdS4_exclusion}, we have graphed the constraints on the coefficients that arise from demanding that the entropy shift is positive. We have included both the constraints from the extremal entropy shift and from considering the shift of all stable black holes. Considering only extremal black holes may be interesting because it is equivalent to the condition that the extremality shift, $\Delta (M / Q)$, is negative. Thus we may look at the constraints implied by positive entropy shift and by negative extremality shift independently. Note that we have divided by $c_1$, which we have already proven to be positive. We may write out the all the constraints obtained:
\begin{align}
    \begin{split}
        &c_1 \ \geq \ 0, \\
        &c_2 \ \geq \ 0, \\
        &c_3 \ \geq \ -\frac{1}{8}c_1 (2 + c_2).
    \end{split}
\end{align}

We have computed the corresponding bounds for AdS${}_5$ through AdS${}_7$. The results may be found in appendix \ref{app:entshift}. We would, however, like to comment on AdS${}_5$, where the positivity of the coefficient of the Riemann-squared term is of  particular interest. The stability analysis yields results that are qualitatively similar to (\ref{firstcondition}) and (\ref{secondcondition}), but with the following definitions
\begin{align}
    \xi^* = 1 - \sqrt{\frac{1 - 2 \nu^2}{1 + 2 \nu^2}}, \qquad \nu^* = \frac{1}{\sqrt{2}}\,.
\end{align}
Once again, we see that large black holes are stable for all values of the charge. 

When we examine the entropy shift in the neutral limit, we find
\begin{align}
    \frac{\pi l^2 }{32 T} c_1 \left( 87 + 164 \nu^2 + 52 \nu^4 \right) \, ,
\end{align}
whose overall sign is completely determined by that of $c_1$.
This means that there are stable black holes where the sign of the entropy shift is the same as the sign of the coefficient of $R_{abcd}^2$. Thus, a positive entropy shift for stable black holes implies that $c_1$ is positive. In fact, a positive value of $c_1$ was the necessary ingredient in \cite{Kats:2007mq} for obtaining the violation of the KSS bound\footnote{We have checked the calculation with a different basis, choosing to use Gauss-Bonnet instead of Riemann squared. As expected, we find that the coefficient of the Gauss-Bonnet term is positive.}. It is also interesting to note that in $d>3$, this sign constraint was shown to follow from the assumption of a unitary tree-level UV completion \cite{Cheung:2016wjt}. The entropy constraints given in this paper are then strictly stronger since they also apply in $d=3$. 

In closing, we stress that we are not claiming that the entropy shift should be universally positive; the proof outlined above only applies when the higher-derivative corrections are generated by integrating out massive fields at tree-level (and relies on assuming that the corresponding solutions minimize the effective action). However, it is interesting that the conjecture that the entropy shift is universally positive appears to suggest that violations of the KSS bound are required to occur.
Our results extend and make more precise the earlier claim by some of us \cite{Cremonini:2009ih} of a link between the WGC and the violation of the KSS bound. We will come back to this point in section \ref{sec:discussion}.

\subsection{Flat Space Limit}

\begin{figure}
    \includegraphics[scale=0.5]{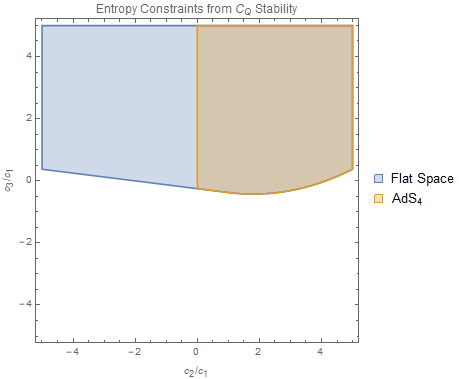}
    \includegraphics[scale=0.5]{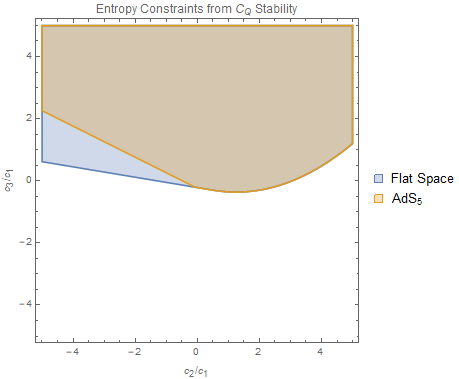}
    \caption{The blue regions are allowed in flat space and the orange in AdS-- note that the AdS regions are a subset of those from flat space.}
    \label{fig: flat space} 
\end{figure}

\noindent As we have pointed out above, we can not compare the results we have given above to the flat space limit. This is because if we impose both $C_Q > 0$ and $\epsilon_T> 0$, we find that there are no stable black holes in the flat space limit $\nu \rightarrow 0$ (as suggested by figure \ref{fig: stability}). In AdS/CFT, we expect that both conditions are necessary to ensure thermodynamic stability; nonetheless, we may remove the condition $\epsilon_T> 0$ in order to compare with the flat space limit. In this case, we find that stability requires
\begin{align}
    \xi^* = 1 - \frac{1}{\sqrt{3}} \sqrt{\frac{1 - 3 \nu^2}{ 1 + 3 \nu^2}}, \qquad \nu^* = \frac{1}{\sqrt{3}},
\end{align}
for the AdS${}_4$ black holes, and 
\begin{align}
    \xi^* = 1 - \frac{1}{\sqrt{2}} \sqrt{\frac{1 - 2 \nu^2}{ 1 + 2 \nu^2}}, \qquad \nu^* = \frac{1}{\sqrt{2}},
\end{align}
for the AdS${}_5$ black holes. This allows for a more direct comparison between the two cases. In figure \ref{fig: flat space}, we contrast the bounds obtained in AdS and flat space. The bounds in AdS are stronger, as they should be given that there is an extra parameter's worth of stable black holes. Note also that $c_1> 0$ is implied by positivity in AdS, but not in flat space, because in flat space there are no stable neutral black holes.  

\section{Discussion}
\label{sec:discussion}

\noindent In this paper, we have examined the relationship between the higher-derivative corrections to entropy and extremality in Anti-de Sitter space. As we have seen, extremality is considerably more complicated in AdS because the relationship between mass, charge, and horizon radius at extremality is non-linear. Nonetheless, we have verified the relation \cite{Cheung:2018cwt, Goon:2019faz} between the entropy shift at fixed charge and mass and the extremality shift at fixed charge and temperature. There is a sharp dependence on which quantities are held fixed in AdS. This is in contrast to flat space, where the linear relationship between mass, charge, and horizon radius removes this issue. We have also provided a more general proof of this relation in appendix \ref{app:goonpenco}, and extended the result to show that there is a third proportional quantity, which is the extremality shift at fixed mass and temperature.

When viewed geometrically, these statements seem almost accidental. In section \ref{sec:euc}, we performed the same calculation from a thermodynamic point of view by computing the free energy from the renormalized on-shell action. From this point of view, issues concerning ``which quantity is held fixed" translate to ``which ensemble is used." In addition to providing an additional check on the results from section \ref{sec:mce}, this provides a non-trivial confirmation of the results of \cite{Reall:2019sah}, which states that the shifted geometry is not needed to compute the thermodynamic quantities. 

Assuming that the entropy shift is positive places constraints on the Wilson coefficients. However, a crucial difference appears in AdS when compared to flat space. The stability criterion depends on the horizon radius over the AdS length, and goes to zero at large horizon radius. This means that there are stable neutral black holes that are asymptotically AdS. For neutral black holes, the entropy shift is dominated by $c_1$, which is the coefficient of the Riemann squared term, so the positivity of the entropy shift implies the positivity of this coefficient. In AdS${}_5$, this coefficient may be related to the central charges of the dual field theory \cite{Henningson:1998gx, Nojiri:1999mh, Blau:1999vz} by 
\begin{align}
    c_1 = \frac{1}{8} \frac{c - a}{c} \, .
\end{align}
Thus, the positivity of the entropy shift appears to be violated in theories where $c - a < 0$. In \cite{Buchel:2008vz}, a number of superconformal field theories were examined, and all were found to satisfy $c - a > 0$. It is worth noting there are non-interacting theories where $c - a< 0$; for example, $\frac{a}{c} = \frac{31}{18}$ for a free theory of only vector fields \cite{Hofman:2008ar}. However, such theories do not have weakly curved gravity duals \footnote{In fact, there are holographic theories with $c<a$. These are the $\mathcal{T}_N$ theories, which arise as $M_5$ branes wrapping punctured Riemann surfaces. As these theories enjoy $\mathcal{N} = 2$ supersymmetry on the boundary, their bulk duals necessarily include massless scalars in the graviton multiplet. Therefore our analysis does not include this case, and it may be interesting to try to extend our work to include scalars in the massless spectrum. We thank Eric Perlmutter for making us aware of this interesting example.}.

The question of whether holographic theories with gravity and gauge fields necessarily correspond to $c-a$ non-negative is interesting for a number of reasons -- both from a fundamental point of view and for phenomenological applications.

In particular, recall that the range of the Wilson coefficients and the sign of $c-a$ played an important role in the physics of the shear viscosity to entropy ratio $\eta/s$ and how it deviates from its universal    $1/4\pi$ result \cite{Policastro:2001yc,Buchel:2003tz}, as discussed extensively in the literature (see \cite{Cremonini:2011iq} for a review of the status of the shear viscosity to entropy bound).
Indeed, it is  interesting to compare our results to the higher-derivative corrections to $\eta/s$, which (for the $\text{AdS}_5$ case of interest to us here) were shown \cite{Myers:2009ij} to be given by 
\begin{align}
    \frac{\eta}{s} = \frac{1}{4 \pi} \left( 1 - 8 c_1 + 4 (c_1 + c_2) \frac{q^2}{r_0^6} \right) \, ,
\end{align}
where $r_0$ is a parameter of the solution defined in \cite{Myers:2009ij}; the factor ${q^2}/{r_0^6}$ goes from 0 (for neutral black holes) to 2 (at extremality). Our bounds on $c_1$ imply that neutral black holes will 
necessarily have a negative viscosity shift, violating the KSS bound. Models where this is realized are known to exist---the first UV complete counter-example to the KSS bound was given in \cite{Kats:2006xp}. 
For extremal black holes, the dependence on $c_1$ drops out and only the sign of $c_2$ matters, ${\eta}/{s} = \frac{1}{4 \pi} \left( 1 + 8 c_2\right)$. For AdS${}_5$, the $c_2$ coefficient may have both positive and negative values. However, imposing the null energy condition implies an additional constraint on the range of $c_2$, which in $\text{AdS}_5$ takes the form
\begin{align}
    \frac{13}{12} \, c_1 + c_2>0 \, .
    \label{eq:NECbound}
\end{align}
This may be seen by first noticing that the definition of the parameter $\gamma$ in equation (\ref{eq:gamma}) implies $\gamma > 0$ as long as the null energy condition holds. Then the bound in (\ref{eq:NECbound}) may be derived from the specific form of $\gamma$ given in (\ref{eq: gammaVal}). This alone is sufficient to bound $c_2$ from below, when $c_1$ is non-negative. 
Thus, one can see that utilizing such constraints it is at least in principle possible to bound $\eta/s$ \emph{from below}, in specific cases. 
To what extent this can be done generically is still an open question.

It might be interesting to try to relate the extremality bounds to the  transport coefficients of the boundary theory in a more concrete way. As the corrections to $\eta / s$ depend only on $c_1$ and $c_2$ in five dimensions, it is clear that the shift to extremality is not captured by the physics that controls $\eta / s$ alone. One might wonder, however, if some other linear combination of transport coefficients, such as the conductivity or susceptibility\footnote{These have been considered in \cite{Kovtun:2008kx}, which already in 2008 had an interesting comment about a possible relation to the WGC.}, might be related to the extremality shift. From a purely CFT point of view, this is certainly not that strange; the philosophy of conformal hydrodynamics is that scaling symmetry ties together ultraviolet quantities ($a, c$) that characterize the CFT to the transport coefficients, which characterize the IR, long-wavelength behavior of the theory. If we believe that EFT coefficients in the bulk are related to these UV quantities (as is known in the case of $c_1$), then a correspondence between higher-derivatives and hydrodynamics is very natural. 
The question is to what extent this can be used to efficiently constrain IR quantities. 
Finally, we should note that extending our analysis to holographic theories that couple gravity to scalars would be useful to make contact with the efforts to generate non-trivial temperature dependence for $\eta/s$ (see \emph{e.g.} the discussion in \cite{Cremonini:2011ej, Cremonini:2012ny}), which is expected to 
play a key role in understanding the dynamics of the strongly coupled quark gluon plasma. 

Our results also have potential to make contact with the work on CFTs at large global charge \cite{Hellerman:2015nra}. As we've seen above, the extremality curve for AdS-Reissner-Nordstr{\"o}m black holes is non-linear even at the two-derivative level. In an analysis of the minimum scaling dimension for highly charged 3D CFTs states of a given charge, it was found \cite{Loukas:2018zjh} that $\Delta \sim q^{3/2}$. This is in striking agreement with the extremality relationship $m \sim q^{3/2}$ that holds for large black holes. The large charge OPE may be powerful because it offers an expansion parameter, $1 / q$, which may be used even for CFTs which are strongly coupled. In principle, it should be possible to match our higher-derivative corrections to the extremality bound with corrections to the minimum scaling dimension that are subleading in $1 / q$. This might allow one to use the large charge OPE to compute the EFT coefficients of the bulk dual of specific theories where the minimum scaling dimensions are known.

\subsection{Weak Gravity Conjecture in AdS}

\noindent One of the motivations for this work is to address to question of to what extent the WGC is constraining in Anti-de Sitter space.
It is not obvious that it should be. In flat space, one looks for higher-derivative corrections to shift the extremality bound $m(q)$ to have a slope that is greater than one. In that case, a single nearly extremal black holes is (kinematically) allowed to decay to two smaller black holes, which can fly apart off to infinity and decay further if they wish.

In AdS, the extremality bound $m(q)$ has a slope that is greater than one at the two-derivative level. Therefore one might expect that large black holes are already able to decay without any new particles or higher-derivative corrections. This picture may be too naive, however; the AdS radius introduces a long range potential that is proportional to $\frac{r^2}{l^2}$. This causes all massive states emitted from the black hole to fall back in, contrary to the situation in flat space. 

A different decay path is provide by the dynamical instability \cite{Gubser:2008px, Hartnoll:2008kx, Hartnoll:2008vx, Denef:2009tp}, whereby charged black branes are unstable to formation of a scalar condensate. This occurs only if the theory also has a scalar with charge $q$ and dimension $\Delta$ that satisfies
\begin{align}
    (m_{\phi} l)^2 \leq \frac{1}{2}(q_{\phi}g M_{Pl} l)^2 - \frac{3}{2} \,.
\end{align}
Note that, even in the limit of large AdS-radius $l$, this does not approach the bound we have for small black holes, which is $m \leq q$. 
Numerical work in \cite{Hartnoll:2008kx} suggests that the endpoint of the instability is a state where all the charge is carried by the scalar condensate. 
Similar requirements appear for the superradiant instability of small black holes \cite{Bhattacharyya:2010yg, Dias:2010ma}. For a more thorough review, see \cite{Nakayama:2015hga}. In either case, it is curious that in AdS, a condition similar to the flat space WGC allows for black holes to decay through an entirely different mechanism.

Another remarkable hint of the WGC comes from its connection to cosmic censorship. In  \cite{Crisford:2017gsb, Horowitz:2019eum}, it is shown that a class of solutions of Einstein-Maxwell theory in AdS${}_4$ that appear to violate cosmic censorship \cite{Horowitz:2016ezu} are removed if the theory is modified to include a scalar whose charge is great enough to satisfy the weak gravity bound\footnote{The bound they consider is the bound for superradiance of small black holes, which requires $\Delta \leq q l$.}.

It may be possible to study these solutions in the presence of higher-derivative corrections. One might ask whether there is a choice of higher-derivative terms such that the singular solutions are removed. It would be interesting to check if this occurs when the higher-derivative terms are those that are obtained by integrating out a scalar of sufficient charge. It would also be interesting to compare constraints obtained by requiring cosmic censorship with constraints due to positivity of the entropy shift.

A more general proof of the WGC in AdS was given in \cite{Montero:2018fns}. In that paper, it was shown that, under mild assumptions, entanglement entropy for the boundary dual of an extremal black brane should go like the surface area of the entangling subregion, which is in tension with the volume law scaling predicted by the Ryu-Takayanagi formula. The contradiction is removed when one introduces a WGC-satisfying state. This violates one of the assumptions that imply the area law for the entropy-- that is, the assumption that correlations decay exponentially with distance.

This form of the WGC in particularly interesting to us because it makes no reference to whether or not the WGC-satisfying state is a particle, or a non-perturbative object like a black hole. Therefore, the contradiction pointed out in that paper may be lifted if the higher-derivative corrections allow for black holes with charge greater than mass. Heavy black holes in AdS have masses far greater than their charge-- therefore we expect that the WGC-satisfying states might be provided by small black holes whose higher-derivative corrections shift the extremality bound to allow slightly more charge.

\section*{Acknowledgements} 

\noindent It is a pleasure to thank Anthony Charles, Cliff Cheung, Simeon Hellerman, Finn Larsen, Grant Remmen, and Gary Shiu for useful conversations and input on this project. This work was supported in part by the U.S.~Department of Energy under grant DE-SC0007859.  S.C. is supported in part by the National Science Foundation Grant PHY-1915038. CRTJ was supported in part by a Leinweber Student Fellowship and in part by a Rackham Predoctoral Fellowship from the University of Michigan.

\appendix

\section{Entropy Shifts from the On-Shell Action}
\label{app:entshift}

\noindent In section \ref{sec:posent}, we computed the constraints on the coefficients in AdS${}_4$. Here we will present the results of this calculation for AdS${}_5$ through AdS${}_7$ using the entropies computed in section \ref{sec:mce}, which corresponds to working in the zero Casimir energy scheme.  For completeness, we also present the Casimir energies for AdS$_5$ and AdS$_7$ that show up in the thermodynamic energy of section \ref{sec:euc} when using a minimal set of counterterms.

\subsection{\texorpdfstring{AdS${}_5$}{AdS5}}

\noindent In AdS${}_5$ we find that the stability condition obtained by demanding positive specific heat and permittivity is given by $\xi < \xi^*$ for $\nu<\nu^*$, with
\begin{align}
    \xi^* = 1 - \sqrt{\frac{1 - 2 \nu^2}{1 + 2 \nu^2}}, \qquad \nu^* = \frac{1}{\sqrt{2}} \, ,
\end{align}
and that all black holes with $\nu>\nu^*$ are stable for all values of the charge. The full entropy shift is simpler to express as a function of charge $q$ than extremality parameter $\xi$. We find 
\begin{align}
\label{dSde5}
    \begin{split}
      &\left( \frac{\partial S }{\partial \epsilon} \right)_{M, Q} \ = \ \frac{\pi}{256 l^6 \nu^8 T} \Big( c_1 \left( 43 q^4 - 24 l^4 q^2 \nu^4 (8 + 5 \nu^2) + 32 l^8 \nu^8(18 + 41 \nu^2 + 13 \nu^4) \right) \\
      & \kern11.5em + 24 c_2 q^2 \left( 3 q^2 - 8 l^4 \nu^4  \right) +72 ( 2 c_3 + c_4)q^4 \Big)\,.
    \end{split}
\end{align}
Note that holographic renormalization in AdS$_5$ with a Riemann-squared correction yields a Casimir energy
\begin{equation}
    E_c=\fft{\omega_3}{16\pi}\left(\fft34l^2-\fft{15}4c_1l^2\right),
\end{equation}
where $\omega_3=2\pi^2$.  This Casimir energy must be removed from the thermodynamic energy in order to obtain the mass $M$ of the black hole.  Alternatively, it can be cancelled right from the beginning by adding an appropriate finite counterterm to the action, in which case the thermodynamic energy would then correspond directly to the mass.  If the Casimir energy is not removed, then the thermodynamic energy shift becomes a combination of mass shift and Casimir energy shift since $E_c$ depends explicitly on the $c_1$ Wilson coefficient.

We find the following expression for the extremal limit,
\begin{align}
    &\left( \frac{\partial S }{\partial \epsilon} \right)_{M, Q} \nonumber\\
    & =\frac{\pi l^2 }{16 T} \left( c_1 (31 + 128 \nu^2 + 138 \nu^4) + 24 c_2 (1 + 2 \nu^2)(1 + 6 \nu^2) + 72(2c_3 + c_4) (1 + 2 \nu^2)^2 \right)  ,
\end{align}
while in the neutral limit we have
\begin{align}
    \left( \frac{\partial S }{\partial \epsilon} \right)_{M, Q} \ = \ \frac{\pi l^2 }{16 T} c_1 \left( 18 + 41 \nu^2 + 13 \nu^4 \right) \, .
\end{align}
Once again, the entropy shift is proportional to $c_1$ in this limit.
\begin{figure}
    \includegraphics[scale=0.59]{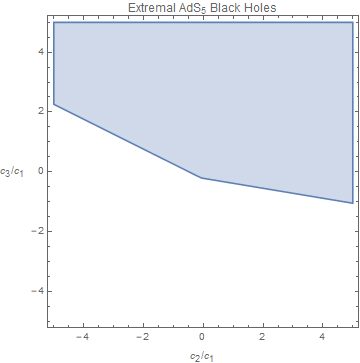}
    \includegraphics[scale=0.59]{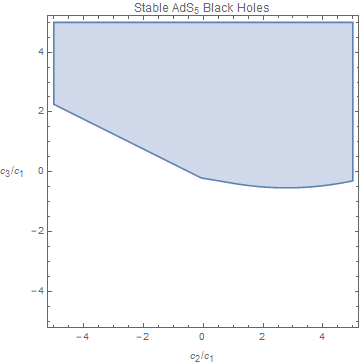}
    \caption{Allowed regions for AdS${}_5$ EFT coefficients.}
    \label{AdS5_exclusion} 
\end{figure}
It is interesting that we do not find a positivity constraint on $c_2$, as we did in AdS${}_4$. There is a lower bound on $c_3 / c_1$ of about -0.5339. The general constraints obtained by the Reduce function of Mathematica are extremely complicated and probably of little interest.

\subsection{\texorpdfstring{AdS${}_6$}{AdS6}}

\begin{figure}
    \includegraphics[scale=0.59]{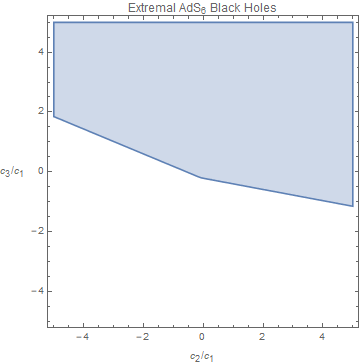}
    \includegraphics[scale=0.59]{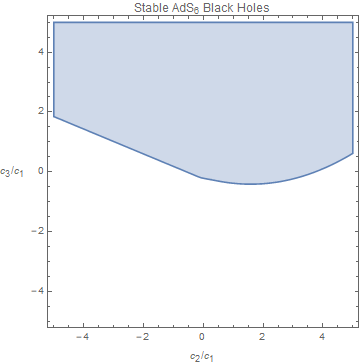}
    \caption{Allowed regions for AdS${}_6$ EFT coefficients.}
    \label{AdS6_exclusion} 
\end{figure}

\noindent In AdS${}_6$ the stability condition obtained by demanding positive specific heat and permittivity is of the same general structure as in AdS${}_5$, 
but with the following identifications:
\begin{align}
    \xi^* = 1 - \sqrt{\frac{3 - 5 \nu^2}{3 + 5 \nu^2}}, \qquad \nu^* = \sqrt{\frac{3}{5}}\,.
\end{align}
The entropy shift is given by:
\begin{align}
    \begin{split}
      &\left( \frac{\partial S }{\partial \epsilon} \right)_{M, Q} \ = \ \frac{\pi}{264 l^9 \nu^{11} T} \Big( c_1 \left( 189 q^4 - 22 l^6 q^2 \nu^6 (36 + 29 \nu^2) + 264 l^{12} \nu^{12} (8 + 17 \nu^2 + 7 \nu^4) \right) \\
      & \kern12em + 2 c_2 q^2 \left( 153 q^2 - 44 l^6 \nu^6 (9 + 5 \nu^2) \right) + 288 ( 2 c_3 + c_4)q^4 \Big) \, ,
    \end{split}
\end{align}
and in the extremal limit takes the form:
\begin{align}
    &\left( \frac{\partial S }{\partial \epsilon} \right)_{M, Q}
    \ = \ \frac{2 \nu \pi l^3 }{99 T} \Bigl( c_1 (369 + 1263 \nu^2 + 1124 \nu^4) + 4 c_2 (3 + 5 \nu^2)(27 + 100 \nu^2) \nonumber\\
    &\kern10em+ 96 (2c_3 + c_4) (3 + 5 \nu^2)^2 \Bigr) \, .
\end{align}
Finally, in the neutral limit we find
\begin{align}
    \left( \frac{\partial S }{\partial \epsilon} \right)_{M, Q} \ = \ \frac{ \nu \pi l^3 }{ T} c_1 \left( 8 + 17 \nu^2 + 7 \nu^4 \right) \, .
\end{align}
Note that no Casimir energy subtraction is needed in AdS$_6$.
We again find that $c_1$ is positive. The other bounds are displayed in figure \ref{AdS6_exclusion}. In AdS${}_6$ and AdS${}_7$, the Reduce function of Mathematica was not able to find the general constraints over all stable values of $\xi$ and $\nu$. However, we believe that the strongest constraints will come from the boundaries of the region of stable black holes. Specifically, we imposed positivity at the neutral $\xi \rightarrow 1$ limit, the extremal $\xi \rightarrow 0$ limit, the planar limit $\nu \rightarrow \infty$ limit, and at $\xi = \xi^*$. We believe this method should give the same answer, and we have checked explicitly that it does in the case for AdS${}_4$ and AdS${}_5$.

\subsection{\texorpdfstring{AdS${}_7$}{AdS7}}

\noindent In AdS${}_7$ the stability window is determined by 
\begin{align}
    \xi^* = 1 - \sqrt{\frac{2 - 3 \nu^2}{2 + 3 \nu^2}}, \qquad \nu^* = \sqrt{\frac{2}{3}} \, ,
\end{align}
and the entropy shift is:
\begin{align}
    \begin{split}
      &\left( \frac{\partial S }{\partial \epsilon} \right)_{M, Q} \\
      &\qquad= \ \frac{\pi^2}{896 l^{12} \nu^{14} T} \Big( c_1 \left( 556 q^4 - 14 q^2 l^8 \nu^8 (160 + 141 \nu^2) + 56 l^{16} \nu^{16} ( 100 + 207 \nu^2 + 8 \nu^4) \right)  \\
      & \kern9em + 80 c_2 q^2 \left( 11 q^2 - 7 l^8 \nu^8 (4 + 3 \nu^2) \right) + 800 ( 2 c_3 + c_4) q^4 \Big)\,.
    \end{split}
\end{align}
The Casimir energy that must be removed from the thermodynamic energy in AdS$_7$ is
\begin{equation}
    E_c=\fft{\omega_5}{16\pi}\left(-\fft58l^4+\fft{35}8c_1l^4\right),
\end{equation}
where $\omega_5=\pi^3$.

\begin{figure}
    \includegraphics[scale=0.59]{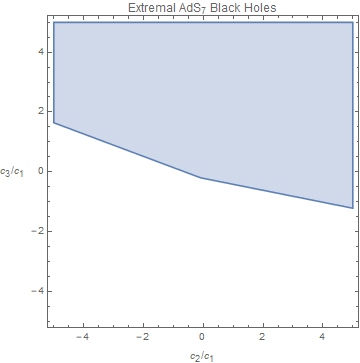}
    \includegraphics[scale=0.59]{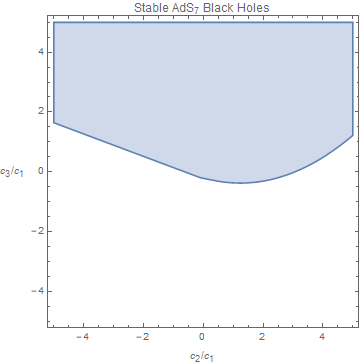}
    \caption{Allowed regions for AdS${}_7$ EFT coefficients.}
    \label{AdS7_exclusion} 
\end{figure}

We find the following expression for the extremal limit,
\begin{align}
\begin{split}
        & \left( \frac{\partial S }{\partial \epsilon} \right)_{M, Q} \ = \ \frac{\pi^2 \nu^2 l^4}{224 T} \Big( c_1  \left( 1384 + 4236 \nu^2 + 3345 \nu^4 \right) \\ 
    &\kern8em + 40 c_2 (2 + 3 \nu^2)(16 + 45 \nu^2) + 800 (2c_3 + c_4)(2 + 3 \nu^2)^2 \Big)\, ,
\end{split}
\end{align}
while in  the neutral limit we find
\begin{align}
    \left( \frac{\partial S }{\partial \epsilon} \right)_{M, Q} \ = \ \frac{ \pi^2 l^2 \nu^2 }{16 T} c_1 \left( 100 + 207 \nu^2 + 93 \nu^4 \right) \, .
\end{align}
Once again, $c_1$ is positive. The other bounds are displayed in figure \ref{AdS7_exclusion}. Again, we used the method of extremizing over the boundaries of the space of stable black holes. 

\section{Another Proof of the Entropy-Extremality Relation} 
\label{app:goonpenco}

\noindent Recent work \cite{Cheung:2018cwt, Goon:2019faz} suggests a remarkable universal relationship between the corrections to extremality and corrections to entropy. Here we will present a simple derivation of this relation using standard thermodynamic identities, including a slight generalization of the relation away from extremality. The statement itself is not specific to black holes, and is in fact a relatively universal statement about infinitesimal deformations of thermodynamic systems. As we explain in detail both here and in appendix \ref{app:vnearext}, the relation we obtain is \textit{formally} correct, but has a subtle physical interpretation.

Consider a thermodynamic system, let $E$ be the total thermal energy, $T$ the temperature, $S$ the entropy and $X$ collectively label a set of extensive thermodynamic variables (for black holes this could be the charge $Q$ and spin $J$). Now consider a small deformation of this system parametrized by a continuous parameter $\epsilon$. The key assumption we will make about this deformation is that it preserves the third law of thermodynamics in the form
\begin{equation}\label{third}
  \lim_{T\rightarrow 0}TS(T,X,\epsilon) = 0,
\end{equation}
for all $\epsilon$ on an open neighbourhood of $\epsilon=0$. We begin with the first law of thermodynamics in the form
\begin{equation}\label{first}
  1 = T\left(\frac{\partial S}{\partial E}\right)_{X,\epsilon}.
\end{equation}
Making use of the triple product identity
\begin{equation}
  \left(\frac{\partial S}{\partial E}\right)_{X,\epsilon}\left(\frac{\partial E}{\partial \epsilon}\right)_{X,S}\left(\frac{\partial \epsilon}{\partial S}\right)_{X,E} = -1,
\end{equation}
we have 
\begin{equation}
  \left(\frac{\partial E}{\partial \epsilon}\right)_{X,S} = -T \left(\frac{\partial S}{\partial \epsilon}\right)_{X,E}.
\end{equation}
Formally inverting $S(T,X,\epsilon)$ gives $T(S,X,\epsilon)$. We can use this to write 
\begin{equation}
   \left(\frac{\partial E}{\partial \epsilon}\right)_{X,S} =  \left(\frac{\partial E}{\partial \epsilon}\right)_{X,T} +  \left(\frac{\partial E}{\partial T}\right)_{X,\epsilon} \left(\frac{\partial T}{\partial \epsilon}\right)_{X,S}.
\end{equation}
Combining these
\begin{equation}
  \left(\frac{\partial E}{\partial \epsilon}\right)_{X,T} = -T \left(\frac{\partial S}{\partial \epsilon}\right)_{X,E} - \left(\frac{\partial E}{\partial T}\right)_{X,\epsilon} \left(\frac{\partial T}{\partial \epsilon}\right)_{X,S}.
\end{equation}
Next, we use (\ref{first}) again
\begin{align}
  \left(\frac{\partial E}{\partial \epsilon}\right)_{X,T} &= -T \left(\frac{\partial S}{\partial \epsilon}\right)_{X,E} - \left(\frac{\partial E}{\partial T}\right)_{X,\epsilon} \left(\frac{\partial T}{\partial \epsilon}\right)_{X,S} \nonumber\\
                                                          &=-T \left(\frac{\partial S}{\partial \epsilon}\right)_{X,E} - T\left(\frac{\partial S}{\partial E}\right)_{X,\epsilon}\left(\frac{\partial E}{\partial T}\right)_{X,\epsilon} \left(\frac{\partial T}{\partial \epsilon}\right)_{X,S} \nonumber\\
                                                          &=-T \left(\frac{\partial S}{\partial \epsilon}\right)_{X,E} - T\left(\frac{\partial S}{\partial T}\right)_{X,\epsilon} \left(\frac{\partial T}{\partial \epsilon}\right)_{X,S},
\end{align}
one final application of the triple product identity gives the \textit{generalized Goon-Penco relation}
\begin{equation}\label{result}
   \left(\frac{\partial E}{\partial \epsilon}\right)_{X,T} +T \left(\frac{\partial S}{\partial \epsilon}\right)_{X,E} = T \left(\frac{\partial S}{\partial \epsilon}\right)_{X,T}.
\end{equation}
Next we make use of the assumption that the deformation does not violate the third law of thermodynamics. Taylor expanding (\ref{first}) we have
\begin{equation}
\label{epsilonexp}
  \lim_{T\rightarrow 0} \left[TS(T,X,\epsilon=0) + \epsilon T \left(\frac{\partial S}{\partial \epsilon}\right)_{T,X}\biggr\vert_{\epsilon=0} +\mathcal{O}\left(\epsilon^2\right)\right] = 0.
\end{equation}
By assumption this is true on an open neighbourhood of $\epsilon=0$ and so it must be true order-by-order in the expansion. This gives
\begin{equation}
\label{limits}
  \lim_{T\rightarrow 0} \;\lim_{\epsilon\rightarrow 0}T\left(\frac{\partial S}{\partial \epsilon}\right)_{T,X} = 0 \, .
\end{equation}
Using this together with (\ref{result}) gives the Goon-Penco relation \cite{Goon:2019faz}
\begin{equation}
    \label{goonpenco}
      \boxed{\lim_{T\rightarrow 0}\left[\left(\frac{\partial E}{\partial \epsilon}\right)_{X,T}\biggr\vert_{\epsilon=0} +T \left(\frac{\partial S}{\partial \epsilon}\right)_{X,E}\biggr\vert_{\epsilon=0}\right] = 0.}
\end{equation}
For the specific application to black hole thermodynamics we identify $E$ with the mass $M$ of the black hole, $X$ with the black hole parameters measured at infinity such as charge $Q$ or angular momentum $J$, and $\epsilon$ with a Wilson coefficient of a four-derivative effective operator. 

In section \ref{sec:mce}, we have pointed out that shift in charge at fixed mass is also proportional to the entropy shift and mass shift. This statement can be derived similarly. By the triple product identity, 
\begin{align}
\begin{split}
    \left( \frac{\partial E}{\partial \epsilon} \right)_{X_i, T} = - \left( \frac{\partial X_i}{\partial \epsilon} \right)_{E, T} \left( \frac{\partial E}{\partial X_i} \right)_{\epsilon, T} \, .
\end{split}
\end{align}
This holds for any extensive quantity. Now we choose $X_i = Q$, and we may identify 
\begin{align}
    \left( \frac{\partial E}{\partial X_i} \right)_{\epsilon, T} = \Phi.
\end{align}
So we find 
\begin{align}
\begin{split}
    \left( \frac{\partial E}{\partial \epsilon} \right)_{Q, T} =  -\Phi \left( \frac{\partial Q}{\partial \epsilon} \right)_{E, T}.
\end{split}
\end{align}
For black holes, this means that the shift in charge is related to the shift in mass.  Neither of them is related to the entropy except at extremality. The result of this is that the entropy shift at extremality may be related to the extremality shift at constant charge or at constant mass,
\begin{align}
\label{gpgen}
    \lim_{T\rightarrow 0} \left( \frac{\partial E}{\partial \epsilon} \right)_{Q, T} = - \lim_{T\rightarrow 0}  \Phi \left( \frac{\partial Q}{\partial \epsilon} \right)_{E, T} =  - \lim_{T\rightarrow 0} T \left(\frac{\partial S}{\partial \epsilon}\right)_{Q,E}.
\end{align}

While the derivation in this appendix is formally valid, the consequences of the expansion in (\ref{epsilonexp}) are subtle and require some commentary. Throughout we have made the implicit assumption that the various thermodynamic quantities are differentiable functions of $\epsilon$, and moreover that $TS(T,X,\epsilon)$ is analytic on an open neighbourhood of $\epsilon=0$, permitting the use of the Taylor series. In general this is a valid assumption for $T\neq 0$, but may \textit{fail} to be valid at $T=0$. As a consequence, the limits in (\ref{limits}) fail to commute
\begin{equation}
  \lim_{T\rightarrow 0} \;\lim_{\epsilon\rightarrow 0}T\left(\frac{\partial S}{\partial \epsilon}\right)_{T,X} \neq \lim_{\epsilon\rightarrow 0} \;\lim_{T\rightarrow 0}T\left(\frac{\partial S}{\partial \epsilon}\right)_{T,X} \;,
\end{equation}
making the physical interpretation of the relation problematic. The formally correct Goon-Penco relation (\ref{goonpenco}) and its corollaries (\ref{gpgen}), treats $\epsilon$ as a free parameter in the model that is taken to zero at the end of the calculation. In a physical application of the low-energy EFT (\ref{Action}), which is derived by matching onto a UV completion, the parameter $\epsilon$ takes a fixed finite value, and thermodynamic quantities at extremality are correctly calculated by taking $T\rightarrow 0$ \textit{before} making a small $\epsilon$ expansion. In such a situation, the Goon-Penco relation is only approximately valid in the \textit{near-extremal} regime $T\sim \epsilon$. A correct treatment at extremality for finite $\epsilon$ is described in detail in appendix \ref{app:vnearext}.

\section{Entropy Shifts of Very-Near Extremal Black Holes}
\label{app:vnearext}

The entropy shift in (\ref{eq:GPshift}), which we rewrite as
\begin{equation}
    \left( \Delta  M \right)_{Q, T = 0} =- T_0 \left(  \Delta S \right)_{Q, M}\qquad(T_0>0) \, ,
\label{eq:neshift}
\end{equation}
was derived at \textit{finite} temperature, and as described in appendix \ref{app:goonpenco}, is a valid approximation in the near-extremal regime $T_0\sim \sqrt{\epsilon}$. For black holes with $T_0 \lesssim \sqrt{\epsilon}$, which we will call very-near extremal black holes, the shift in the horizon is not analytic in $\epsilon$ and (\ref{eq:neshift}) is no longer valid. The presence of a double root at extremality implies that the shift to the horizon radius, and therefore to the entropy, is proportional to $\sqrt{\epsilon}$. Nonetheless, a proportionality between the entropy shift and extremality shift still holds. This has been stressed in \cite{Hamada:2018dde, Loges:2019jzs}, where it was shown that, at extremality, the same combination of EFT coefficients that appear in the extremality shift is required to be positive in order to ensure that the entropy shift is real.
In fact, near extremality the form of the entropy-extremality relationship is nearly identical to the one of \cite{Goon:2019faz}, up to an extra factor of $1/2$. The correct relation takes the form
\begin{equation}
\label{VNErelation}
    \left( \Delta  M \right)_{Q, T = 0} =- \frac{1}{2} T \left(  \Delta S \right)_{Q, M}\qquad(T_0=0) \, ,
\end{equation}
where now $T$ is the temperature of the corrected black hole.  In particular, while the leading order temperature vanishes, the corrected solution is evaluated at fixed mass $M$ and charge $Q$, and hence is no longer extremal.

In order to see the transition between (\ref{eq:neshift}) and (\ref{VNErelation}), we may revisit the derivation of the mass and entropy shifts of section~\ref{sec:mce}.  Since the entropy shift is calculated at fixed mass and charge, it is convenient to rewrite the geometry shift in terms of $M$ and $Q$.  In this case, the radial function takes the form
\begin{equation}
    g(r)=g_0(r)+\Delta g=k-\fft{\hat M}{r^{d-2}}+\fft{\hat Q^2}{4r^{2(d-2)}}+\fft{r^2}{l^2}+\epsilon\Delta g,
\end{equation}
where here $\Delta g$ takes into account the higher-derivative corrections to the geometry along with the contribution of the $\rho$ factor to the physical mass $M$ in (\ref{eq:massdef}). Note that, to avoid extra volume factors, we have defined
\begin{equation}
    M=\fft{\omega_{d-1}(d-1)}{16\pi}\hat M,\qquad Q=\fft{\omega_{d-1}}{16\pi}\sqrt{\fft{(d-2)(d-1)}2}\hat Q\,.
\end{equation}
The entropy shift at fixed $(M,Q)$ can then be obtained from the horizon shift according to (\ref{eq:Sshift})
\begin{equation}
    \left(\Delta S\right)_{Q,M}=\fft{\omega_{d-1}(d-1)}4r_0^{d-2}\Delta r_h\,+\cdots,
    \label{eq:esh}
\end{equation}
where $r_0$ is the leading order horizon location, and the ellipses denote the additional higher order corrections to the Wald entropy.  These additional terms will be unimportant in either of the near extremal or very near extremal limits, where the shift to the horizon radius dominates.

Our goal now is to rewrite the horizon shift $\Delta r_h$ in terms of the mass shift and temperature.  For the mass shift, we always consider the shift of the extremal mass at fixed charge, which can be taken from (\ref{mass_shift})
\begin{equation}
    \left(\Delta M\right)_{Q,T=0}=\fft{\omega_{d-1}(d-1)}{16\pi}r_0^{d-2}\epsilon\Delta g\,.
    \label{eq:msh}
\end{equation}
Note that the term in (\ref{mass_shift}) proportional to $\rho$ is absorbed into the shift $\Delta g$ in this expression.  In any case, we see that in order to connect the entropy shift (\ref{eq:esh}) to the mass shift (\ref{eq:msh}), we would like to have a relation between $\Delta r_h$ and $\Delta g$.  The first step here is to use the fact that the radial function $g(r)$ vanishes at the horizon
\begin{equation}
    0=g(r_h)=g_0(r_0+\Delta r_h)+\epsilon\Delta g=g_0(r_0)+\Delta r_hg_0'(r_0)+\ft12(\Delta r_h)^2g_0''(r_0)+\cdots+\epsilon\Delta g\,.
\end{equation}
Note that we have not expanded $\Delta g(r_h)=\Delta g(r_0)+\cdots$ since we already consider it to be a small quantity.  Since $r_0$ is the uncorrected horizon, $g_0(r_0)$ always vanishes.  However, we need to keep the next two terms in the expansion since $g_0'(r_0)$ will vanish for a leading order extremal black hole \cite{Hamada:2018dde}. The horizon shift can then be obtained by solving
\begin{equation}
    \Delta r_hg_0'(r_0)+\ft12(\Delta r_h)^2g_0''(r_0)=-\epsilon\Delta g\,,
    \label{eq:drdg}
\end{equation}
or equivalently
\begin{equation}
     \Delta r_hg_0'(r_0)+\ft12(\Delta r_h)^2g_0''(r_0)=-\fft{16\pi}{\omega_{d-1}(d-1)}r_0^{-d+2}\left(\Delta M\right)_{Q,T=0}\,.
     \label{eq:hosh}
\end{equation}

In principle, we can now insert the horizon shift from (\ref{eq:hosh}) into the expression (\ref{eq:esh}) to obtain a general relation between the entropy shift and the mass shift.  However, as it stands, this result would depend on the derivatives $g_0'(r_0)$ and $g_0''(r_0)$ of the radial function.  To replace this with more physical quantities, we note that the temperature of the corrected black hole can be obtained from $g(r)$ according to
\begin{equation}
    T=\fft1{4\pi}g'(r_h)=\fft1{4\pi}g'(r_0+\Delta r_h)=\fft1{4\pi}\left(g_0'(r_0)+\Delta r_hg_0''(r_0)
    +\cdots+\epsilon\Delta g'\right)\,.
\end{equation}
Although this depends on $\Delta g'$, this term can be ignored as long as we are in the near (or very near) extremal limit since either $g_0'(r_0)$ or $\Delta r_hg_0''(r_0)$ will dominate.  More precisely, $g_0'(r_0)=\mathcal O(\epsilon^0)$ unless it vanishes, and when it does, (\ref{eq:drdg}) along with $g_0''(r_0)=\mathcal O(\epsilon^0)$ demonstrates that $\Delta r_hg_0''(r_0)=\mathcal O(\sqrt\epsilon)$. As a result, we can take
\begin{equation}
    T=\fft1{4\pi}\left(g_0'(r_0)+\Delta r_hg_0''(r_0)\right)\,.
\end{equation}
In the general case, this provides only a single relation between the derivatives $g'(r_0)$ and $\Delta r_hg''(r_0)$ and the temperature $T$.  However, we can also introduce the temperature of the uncorrected black hole, $T_0=g_0'(r_0)/4\pi$, which can be allowed to vanish.  We now solve for the derivatives
\begin{equation}
    g_0'(r_0)=4\pi T_0\,,\qquad \Delta r_hg_0''(r_0)=4\pi(T-T_0)\,.
\end{equation}
Inserting this into (\ref{eq:hosh}) gives us the desired relation between the horizon shift and mass shift
\begin{equation}
    \left(\Delta M\right)_{Q,T=0}=-\fft{\omega_{d-1}(d-1)}4r_0^{d-2}\fft{T_0+T}2\Delta r_h\,,
\end{equation}
which yields
\begin{equation}
    \left(\Delta M\right)_{Q,T=0}=-\fft{T_0+T}2\left(\Delta S\right)_{Q,M}\,,
    \label{eq:finale}
\end{equation}
upon substitution of (\ref{eq:esh}).  Note that this relation is expected to break down away from extremality where the various approximations taken above will no longer be valid.

The relation (\ref{eq:finale}) encompasses both the near and very near extremal cases, (\ref{eq:neshift}) and (\ref{VNErelation}).  When working somewhat away from extremality, we start with $T_0>0$.  In this case, we have the scaling
\begin{equation}
    \Delta r_h\sim\epsilon\,,\qquad\Delta S\sim\epsilon\,,\qquad\Delta M\sim\epsilon\,,
    \label{eq:scaling1}
\end{equation}
along with $T=T_0+\mathcal O(\epsilon)$.  Taking this into account then directly yields (\ref{eq:neshift}).  On the other hand, if we start from extremality, $T_0=0$, we have instead
\begin{equation}
    \Delta r_h\sim\sqrt\epsilon\,,\qquad\Delta S\sim\sqrt\epsilon\,,\qquad\Delta M\sim\epsilon\,,
    \label{eq:scaling2}
\end{equation}
and $T=\mathcal O(\sqrt\epsilon)$, which leads to the extremal case (\ref{VNErelation}).  The transition between these two cases occurs when $\Delta T\sim T$, which more precisely defines the very near extremal limit.

\subsection{Entropy Shift for Heterotic Black Holes}

The above discussion can be applied to an interesting example. Recent work has considered the leading $\alpha'$-corrections to dyonic Reissner-Nordstr{\"o}m black holes embedded in heterotic string theory \cite{Cano:2019oma, Cano:2019ycn}. Though the four-dimensional backgrounds considered in these papers are asymptotically flat, we would like to briefly comment on them in connection with the universal entropy-extremality relationship.

To connect with the above, we would like to consider the corrections to the geometry. These are given \cite{Cano:2019ycn} by the radial function 
\begin{align}
    g(r) = 1 - \frac{2 M}{r} + \frac{p^2}{2 r^2} + \Delta g(r), \qquad \Delta g(r) =  - \alpha' \frac{p^2}{4 r^4} \left[ 1 - \frac{3 M}{2 r} + \frac{11 p^2}{40 r^2}  \right] \, ,
\end{align}
where the constant $p$ denotes the charge in the notation of \cite{Cano:2019ycn}.
With these variables, the leading-order solution is extremal when $M = p/\sqrt{2}$. Using this expression and (\ref{eq:msh}), we can compute the shift in the extremal mass at a given charge $p$,
\begin{align}
    (\Delta M)_{p, T = 0} = \frac{1}{2} r_0 \Delta g(r_0) = - \alpha' \frac{1}{4 (p/\sqrt{2})}\left( 1 - \frac{3}{2} + \frac{22}{40 } \right) = \, - \frac{1}{80} \frac{\alpha'}{ p/\sqrt{2}} \, ,
\end{align}
where we have computed $\Delta g$ at extremality in the last equality. This agrees with \cite{Cano:2019ycn}.

Next, we consider the entropy. In \cite{Cano:2019ycn}, this is shown to be 
\begin{align}
    S = \pi r_h^2 \left\{ 1 + \alpha' \left[ \frac{M}{r_h^3} - \frac{3 p^2}{8 r_h^4} \right] \right\}  .
\end{align}
Just as we have found in the body of the paper, this entropy includes corrections from the higher-derivative terms in the Lagrangian; these are the explicit $\alpha'$ terms that appear in the expression above. It also includes corrections from the horizon radius, which are implicit in the $r_h^2$ that appears in the first term. Thus, we may rewrite the entropy more explicitly as
\begin{align}
    S = \pi r_0^2 \left\{ 1 + 2 \frac{\Delta r_h}{r_0} + \alpha' \left[ \frac{M}{r_0^3} - \frac{3 p^2}{8 r_0^4} \right] \right\} .
\end{align}
The corrections are then given by 
\begin{align}
    (\Delta S)_{M, p} =  2 \pi r_0 \Delta r_h + \alpha' \pi \left[ \frac{M}{r_0} - \frac{3 p^2}{8 r_0^2} \right]  \, .
    \label{eq:cano_ent_cor}
\end{align}
We need the temperature to complete the relation. Recall that 
\begin{align}
    T = \frac{1}{4 \pi} \left( g_0'(r_0) + \Delta r g''_0(r_0) + \cdots \right) \, .
\end{align}
The order of both of these corrections in $\alpha'$ will depend on the order of $\Delta r_h$. Thus, to determine the entropy-extremality relationship we need to compute the correction to the horizon radius, which may be obtained from the geometry shift via equation (\ref{eq:drdg}). This procedure gives different results away from extremality and near extremality. Away from extremality, we have 
\begin{align}
    T \sim T_0 = \frac{1}{4 \pi} g'_0(r_0) \, ,
\end{align}
and the horizon shift is%
\footnote{it is not easy to compare this to (4.12) of \cite{Cano:2019ycn} because they have numerically inverted the expression to ensure that $\Delta r$ is only a function of $M$ and $p$. However, both results for $\Delta r$ lead to the same entropy shift at extremality.}
\begin{align}
    \Delta r = - \frac{\Delta g}{4 \pi T_0} \, .
\end{align}
The corresponding entropy shift becomes
\begin{align}
    (\Delta S)_{M, p} = \frac{1}{T_0} \alpha' \frac{p^2}{8 r_0^3} \left[ 1 - \frac{3 M}{2 r_0} + \frac{11 p^2}{40 r_0^2}  \right] + \alpha' \pi \left[ \frac{M}{r_0} - \frac{3 p^2}{8 r_0^2} \right] \, .
\end{align}
If we then take this expression to extremality, we take the limit where $T_0 \rightarrow 0$. Since the first term dominates in this limit, we may write the entropy shift as 
\begin{align}
    T_0 (\Delta S)_{M, p} =  \frac{1}{80 } \frac{\alpha'}{p / \sqrt{2}} \, ,
\end{align}
where we have made use of the extremal value of the mass.
This gives the correct relation away from extremality,
\begin{align}
    (\Delta M)_{p, T = 0} = - T_0 (\Delta S)_{M, p} \, .
\end{align}

Now let us perform the same analysis in the very-near extremal limit. This is the limit where $T_0 = 4 \pi g_0' \rightarrow 0$, so that for the temperature we have 
\begin{align}
    T = \frac{1}{4 \pi} \Delta r_h g''_0(r_0) 
\end{align}
to leading order in $\alpha'$. In this limit, we now extract a different result from (\ref{eq:drdg}),
\begin{align}
    \Delta r_h = \sqrt{- 2 \frac{\Delta g}{g''_0(r_0)} }\, .
\end{align}
In this case, the first term of (\ref{eq:cano_ent_cor}) still dominates the entropy corrections, but here this is because it is $\mathcal{O}(\sqrt{\alpha'})$, rather than because the denominator diverges. Thus, very-near extremality the entropy shift is 
\begin{align}
    (\Delta S)_{M, p} =  2 \pi r_0 \Delta r_h \, ,
\end{align}
and therefore in this limit we find 
\begin{align}
    T \,(\Delta S)_{M, p} \, = \,\frac{1}{2}\, r_0 \, \Delta r_h^2 \, g''_0(r_0) \, = \, - r_0 \, \Delta g \, =\,   \frac{1}{40} \, \frac{\alpha'}{ p/\sqrt{2}} \, .
\end{align}
This gives the very-near extremal relation:
\begin{align}
    (\Delta M)_{p, T = 0} = -\frac{1}{2} T (\Delta S)_{M, p}  \, .
\end{align}
This agrees with the discussion of \cite{Cano:2019ycn} and confirms their claim that this is the correct form of the relationship in the extremal limit. It also confirms the scaling relations in (\ref{eq:scaling1}) and (\ref{eq:scaling2}). In both cases, $\Delta M$ is order $\alpha'$. However, in the very-near extremal limit we see the non-analytic behavior discussed above, namely that $T \sim \sqrt{\alpha'}$ and $\Delta S \sim \sqrt{\alpha'}$. It is also important to note that in both cases, we took the extremal limit. However, the relative size of $T_0$ and $\alpha'$ dictates the correct order of limits to take. The derivation ``away from extremality" remains valid for $T_0 \geq \sqrt{\alpha'}$. For very small temperatures --- that is, for $T_0 \lesssim \sqrt{\alpha'}$ --- the additional subtleties discussed in this section arise, and we need to use the ``very-near extremal" derivation. Nonetheless, in both cases there is a proportionality between the mass shift and the temperature times the entropy shift. 

\bibliographystyle{JHEP}
\bibliography{cite.bib}

\end{document}